%% file: MAIN.tex
\documentclass[11pt]{article}
\usepackage[a4paper,hmargin=1.0in,vmargin=1.0in]{geometry}
\usepackage{mathtools,amsthm,amssymb,setspace,sectsty,xcolor}
\usepackage[round]{natbib}
\usepackage[linktocpage=true,pagebackref=true,colorlinks,allcolors=blue,bookmarks=true,bookmarksopen,bookmarksnumbered]{hyperref}

\newtheoremstyle{DStheorem}
  {\topsep}
  {\topsep}
  {\itshape}
  {0pt}
  {\scshape}
  {.}
  { }
  {\thmname{#1}\thmnumber{ #2}\thmnote{ (#3)}}
\theoremstyle{DStheorem}
\newtheorem{theorem}{Theorem}[section]
\newtheorem{lemma}[theorem]{Lemma}
\newtheorem{claim}[theorem]{Claim}

\let\oldproofname=\proofname
\renewcommand{\proofname}{\rm\sc{\oldproofname}}

\newcommand{\bs}[1]{\boldsymbol{#1}}
\newcommand{\bbR}{\mathbb{R}}

\newcommand{\bbZ}{\mathbb{Z}}
\newcommand{\eps}{\epsilon}
\newcommand{\opt}{\mathrm{OPT}}
\newcommand{\apx}{\mathrm{APX}}
\newcommand{\topt}{\widetilde{\opt}}
\newcommand{\ex}[1]{\mathbb{E}\left[#1\right]}
\newcommand{\expar}[1]{\mathbb{E}[#1]}
\newcommand{\pr}[1]{\mathrm{Pr}\left[#1\right]}
\newcommand{\prpar}[1]{\mathrm{Pr}[#1]}

\newcommand{\lcm}{\mathrm{LCM}}
\newcommand{\mydisc}{\mathrm{disc}}
\newcommand{\mycont}{\mathrm{cont}}
\newcommand{\mymod}{\mathrm{mod}}
\newcommand{\mytower}{\mathrm{tower}}

\sectionfont{\large} \subsectionfont{\normalsize}
\allowdisplaybreaks
\makeindex

\begin{document}

\begin{titlepage}

\pagenumbering{Roman}

\title{New Approximation Guarantees for \\
The Inventory Staggering Problem}
\author{%
Noga Alon\thanks{Department of Mathematics, Princeton University, Princeton, NJ, 08544. Email: {\tt nalon@math.princeton.edu}. 
Research supported in part by NSF grant DMS-2154082.}%
\and%
Danny Segev\thanks{School of Mathematical Sciences and Coller School of Management, Tel Aviv University, Tel Aviv 69978, Israel. Email: {\tt segevdanny@tauex.tau.ac.il}. Research supported by Israel Science Foundation grant 1407/20.}}
\date{}
\maketitle

\begin{abstract}
Since its inception in the mid-60s, the inventory staggering problem has been explored and exploited in a wide range of application domains, such as production planning, stock control systems, warehousing, and aerospace/defense logistics. However, even with a rich history of academic focus, we are still very much in the dark when it comes to cornerstone computational questions around inventory staggering and to related structural characterizations, with our methodological toolbox being severely under-stocked.

The central contribution of this paper consists in devising a host of algorithmic techniques and analytical ideas --- some being entirely novel and some  leveraging well-studied concepts  in combinatorics and number theory --- for surpassing essentially all  known approximation guarantees for the inventory staggering problem. In particular, our work demonstrates that numerous structural properties open the door for designing polynomial-time approximation schemes, including polynomially-bounded cycle lengths, constantly-many distinct time intervals, so-called nested instances, and pairwise coprime  settings. These findings offer substantial improvements over currently available constant-factor approximations and resolve outstanding open questions in their respective contexts \citep{TeoOT98, HumST05, HochbaumR19, HochbaumR20}. In parallel, we develop new theory around a number of yet-uncharted questions, related to the sampling complexity of peak inventory estimation as well as to the plausibility of groupwise synchronization. Interestingly, we establish the global nature of inventory staggering, proving that there are $n$-item instances where, for every subset of roughly $\sqrt{n}$ items, no policy improves on the worst-possible one by a factor greater than $1+\eps$, whereas for the entire instance, there exists a policy that outperforms the worst-possible one by a factor of nearly $2$, which is optimal.
\end{abstract}


\end{titlepage}

\setcounter{page}{2}
\tableofcontents

\newpage
\pagestyle{plain}
\pagenumbering{arabic}
\setcounter{page}{1}
\input{TEX-Intro}
\input{TEX-Impossibility}
\input{TEX-Approx-Cycle-Length}
\input{TEX-Approx-Distinct-Intervals}
\input{TEX-Approx-Nested}

\input{TEX-Approx-Coprime}
\input{TEX-Conclusions}

\addcontentsline{toc}{section}{Bibliography}
\bibliographystyle{plainnat}
\bibliography{BIB-Staggering}


\end{document}

%% file: TEX-Intro.tex
\section{Introduction}

The principal goal of this paper is to deepen our understanding and to inject fresh algorithmic ideas into a foundational inventory management model, commonly known as the inventory staggering problem. Since its inception in the mid-60s \citep{Homer1966, PageP76,  HartleyT82}, this paradigm has been explored and exploited in a wide range of application domains, such as  production planning, stock control systems, warehousing, and aerospace/defense logistics, just to name a few. Concurrently, on the theoretical front, we have been witnessing a steady stream of advances in this context, including an appreciable number of highly innovative papers; see, e.g., \citet{Zoller77}, \citet{GallegoSS92}, \citet{TeoOT98}, \citet{HumST05}, and \citet{HochbaumR19}. That said, even with a rich history of academic focus, we are still very much in the dark when it comes to cornerstone computational questions around inventory staggering and to related structural characterizations, with our methodological toolbox being severely under-stocked.

Broadly speaking, while inventory staggering problems present themselves in various configurations, they all fundamentally address the challenge of efficiently synchronizing multiple periodic replenishment policies via horizontal offsets, with the objective of minimizing peak storage requirement across a given planning horizon. In spite of its deceivingly-simple geometric interpretation, what renders this synchronization task particularly elusive is the interplay between numerous replenishment policies, whose individual cycle lengths and ordering quantities may be completely unrelated. Consequently, even very elementary questions, such as evaluating the peak storage requirement of a given solution in polynomial time, remain unresolved to date. Moreover, when we turn the spotlight to optimization-focused questions, nearly all known research directions continue to lack definitive answers. To delve into the finer details of these questions and to preface our main contributions, we proceed by providing a formal mathematical description of the inventory staggering problem in its most general form.

\subsection{Model formulation} \label{subsec:model_definition}

\paragraph{Input description and replenishment policies.} Whether one considers discrete forms of the inventory staggering problem or their continuous counterparts, they are all defined with respect to  a finite collection of $n$ items. Across the continuous planning horizon $[0,\infty)$, the inventory level of each item $i \in [n]$ is replenished via an individual stationary order sizes and stationary intervals (SOSI) policy. Here, $i$-orders of identical quantities are evenly positioned along the timeline, each leaving us with zero inventory upon arriving at its subsequent order. Consequently, any such policy is fully characterized by the next three features: 
\begin{itemize}
    \item The uniform time interval $T_i$ between successive $i$-orders.

    \item The number of units $H_i$ comprising every $i$-order.

    \item The horizontal shift $\tau_i \in [0,T_i]$.
\end{itemize}
It is important to mention that both the time interval $T_i$ between successive $i$-orders and the ordering quantity $H_i$ are assumed to be integer-valued, specified as part of our input description. By contrast, $\tau_i$ is a decision variable, which will be restricted to integer values in the discrete problem setting, and to real values in the continuous setting. Given these parameters, $i$-orders will be placed at the time points
\begin{equation} \label{eqn:seq_orders}
\ldots \quad , \quad -2T_i + \tau_i \quad , \quad -T_i + \tau_i \quad, \quad \tau_i \quad, \quad T_i + \tau_i \quad, \quad 2T_i + \tau_i \quad, \quad \ldots \quad,    
\end{equation}
each consisting of exactly $H_i$ units, meaning that zero-inventory levels are reached at each of these points. For simplicity, this policy will be designated by ${\cal P}^i_{\tau_i}$, noting that we have extended the sequence~\eqref{eqn:seq_orders} into the negative orthant just to avoid convoluted notation later on.

\paragraph{Inventory levels and objective value.} Based on the preceding discussion, it is easy to verify that, when replenished according to ${\cal P}^i_{\tau_i}$, the inventory level of item $i$ at any time $t \in \bbR$ is given by 
\[ I_i( {\tau_i}, t) ~~=~~ H_i \cdot \left( 1 - \frac{ t-  \lfloor t \rfloor_{ {\cal P}^i_{\tau_i} } }{ T_i } \right) \ , \]
where $\lfloor \cdot \rfloor_{ {\cal P}^i_{\tau_i} }$ is an operator that rounds its argument down to the nearest ordering point in ${\cal P}^i_{\tau_i}$. As such, for any shift vector $\tau = (\tau_1, \ldots, \tau_n)$, the overall inventory level of the joint policy ${\cal P}_{\tau} = ({\cal P}^1_{\tau_1}, \ldots, {\cal P}^n_{\tau_n})$ at any time $t \in \bbR$ is precisely $I_{ \Sigma }( \tau, t) = \sum_{i \in [n]} I_i( \tau_i, t)$.

Now, let us observe that, since the time intervals $T_1, \ldots, T_n$ are assumed to be integers, the policy ${\cal P}_{\tau}$ is necessarily cyclic, with a cycle length of $\Lambda = \lcm( T_1, \ldots, T_n)$. In turn, the peak inventory level of this policy can be written as  
\[ I_{ \max }( \tau) ~~=~~ \max_{t \in [0,\Lambda]} I_{ \Sigma }( \tau, t) \ . \]
As a side note, the above-mentioned maximum is indeed attained, since the policy ${\cal P}_{\tau}$ has a finite number of ordering points across $[0,\Lambda]$, and since the function $I_{ \Sigma }( \tau, \cdot)$ is decreasing between any successive pair of these points. 

In the discrete inventory staggering problem, our objective is to determine a shift vector $\tau \in \bbZ^n_+$ for which the peak inventory level $I_{ \max }( \tau)$ of the policy ${\cal P}_{\tau}$ is minimized. Here, the optimum value is clearly attained at an integer time point, and for any given instance ${\cal I}$, this measure  will be designated by 
\[ \opt^{ \mydisc }( {\cal I} ) ~~=~~ \min_{ \tau \in \bbZ^n_+ } \max_{t \in [0,\Lambda] \cap \bbZ} I_{ \Sigma }( \tau, t) \ . \]
In the continuous setting, we arrive at precisely the same formulation, except for expanding our solution space to $\tau \in \bbR^n_+$. The optimum peak inventory level of such instances, which is not necessarily attained at an  integer time point, will be denoted by  
\[ \opt^{ \mycont }( {\cal I} ) ~~=~~ \min_{ \tau \in \bbR^n_+ } \max_{t \in [0,\Lambda]} I_{ \Sigma }( \tau, t) \ . \]

\subsection{Known results and main contributions} \label{subsec:prev_new_results}

The central contribution of this paper consists in devising a host  of algorithmic techniques and analytical ideas --- some being entirely novel and some  leveraging well-studied concepts  in combinatorics and number theory --- for surpassing essentially all  known approximation guarantees for the inventory staggering problem. Concurrently, we develop new theory around a number of yet-uncharted questions, related to the sampling complexity of peak inventory estimation as well as to the plausibility of groupwise synchronization. 

For ease of exposition, we proceed by providing a formal account of our main findings, subsequent to a concise preamble on existing  state-of-the-art results in each of these directions. As a side note, the upcoming presentation order is primarily designed to offer logical and user-friendly  content; this order is uncorrelated with importance, uniqueness, or magnitude of improvement. Additionally, given the extensive body of work dedicated to studying inventory staggering problems, including rigorous methods, heuristics, and experimental evidence, we cannot do justice and exhaustively discuss these research domains. For an in-depth literature review, readers are referred to useful  articles in this context \citep{GallegoQS96, Hall98, MurthyBR03, Boctor10, HochbaumR19} as well as to the references
therein.

\paragraph{Main result 1: Sample complexity of peak evaluation.} As previously mentioned, efficiently evaluating the peak inventory level $I_{ \max }( \tau)$ of a given replenishment policy ${\cal P}_{\tau}$ has remained an unresolved issue since the very inception of inventory staggering, forcing most efforts around this problem to focus on stylized formulations. In Section~\ref{sec:impossible}, we provide the first rigorous evidence for the inherent difficulty of this question, by deriving an exponential lower bound on its sample complexity. Specifically, suppose we randomly draw $M$ independent time points $X_1, \ldots, X_M \sim U \{ 0, \ldots, \Lambda - 1 \}$, and consider the maximal inventory level $\tilde{I}_{ \max }^M = \max_{m \in [M]} I_{ \Sigma }(\tau^*, X_m)$ across these points as our estimator for $I_{ \max }( \tau^*)$, where $\tau^*$ is a given optimal shift vector. Then, as an immediate conclusion of the next theorem, $2^{ \Omega(\eps^2 n ) }$ random points should generally be drawn in order to estimate $I_{ \max }( \tau^*)$ within factor $\frac{ 1 }{ 2 } + \eps$ with constant probability.

\begin{theorem} \label{thm:LB_random_sample}
There exists a discrete inventory staggering instance, together with an optimal shift vector $\tau^* \in \bbZ^n_+$, for which
\[ \pr{ \tilde{I}_{ \max }^M \geq \left( \frac{ 1 }{ 2 } + \eps \right) \cdot I_{ \max }( \tau^* )} ~~\leq~~ M e^{ - \eps^2 n / 6 } \ . \]
\end{theorem}

\paragraph{Main result 2: Infeasibility of groupwise synchronization.} Our next contribution is motivated by the yet-unknown usefulness of pairwise synchronization, recently investigated by \citet{Segev24} in the context of economic warehouse lot scheduling. Deferring the nuts-and-bolts of this concept to Section~\ref{sec:impossible}, at a high level, suppose we initially compute a near-optimal shift vector $\tau^{ij}$ for each pair of items $i \neq j$, hoping to meaningfully drop below the trivial peak inventory level of $H_i + H_j$ for a significant portion of these pairs. Then, the essential question is whether there exists a partition ${\cal M}$ of the item set into pairs, such that $\sum_{(i,j) \in {\cal M}} I_{\max}( \tau^{ij} ) \leq (1- \delta) \cdot \sum_{i \in [n]} H_i$, for some absolute constant $\delta > 0$. As explained later on, such a result would immediately lead to a $2(1-\delta)$-approximation for the discrete inventory staggering problem in its utmost generality, beating the best known approximation guarantee for arbitrarily-structured instances. Surprisingly, we prove that this approach cannot succeed for pairs of items, for triplets, or for subsets of much larger size. Specifically, as a direct corollary of Theorem~\ref{thm:LB_group_sync} below, even when optimal shift vectors are readily available for subsets of nearly $\sqrt{n}$ items, gluing them together across all subsets would still be far from optimal by a factor of $2-O(\eps)$.

\begin{theorem} \label{thm:LB_group_sync}
There exists a discrete inventory staggering instance ${\cal I}$ such that:
\begin{enumerate}
    \item $\opt^{ \mydisc }( {\cal I} ) \leq (\frac{ 1 }{ 2 } + \eps) \cdot \sum_{i \in [n]} H_i$.

    \item $\opt^{ \mydisc }( {\cal I}_{\hat{\cal F}} ) \geq (1 - \eps) \cdot \sum_{i \in \hat{\cal F} } H_i$, for every subset $\hat{\cal F}$ of $O( \frac{ \sqrt{n} }{ \log n } )$ items.    
\end{enumerate}
\end{theorem}

\paragraph{The boundary between trivial and non-trivial.} Moving forward, to clearly mark an informative reference point for recognizing non-trivial approximation guarantees, it is worth keeping in mind that the peak inventory level of any shift vector is always within factor $2$ of optimal. This property is a direct outcome of the simple classic average-space bound (see Lemma~\ref{lem:avg_space_LB}), stating that $\opt^{ \mydisc }( {\cal I} ) \geq \frac{ 1 }{ 2 } \cdot \sum_{i \in [n]} H_i \cdot (1 + \frac{ 1 }{ T_i } )$ and $\opt^{ \mycont }( {\cal I} ) \geq \frac{ 1 }{ 2 } \cdot \sum_{i \in [n]} H_i$.

\paragraph{Main result 3: Approximation scheme in terms of cycle length.} As far as proximity to optimum is concerned, state-of-the-art performance  guarantees for the discrete inventory staggering problem have recently culminated to polynomial-time approximation schemes (PTAS), when the cycle length $\Lambda = \lcm( T_1, \ldots, T_n)$ we are facing is bounded by a constant, i.e., $\Lambda = O(1)$. In particular,  when items are associated with identical time intervals, \citet{HochbaumR19} employed dynamic programming ideas to devise an approximation scheme in $O(n \cdot (\frac{ 1 }{ \eps })^{ O(\Lambda + 1/\eps) } )$ time. Shortly thereafter, \citet{HochbaumR20} were  successful in eliminating the identical-intervals assumption, albeit ending up with an $O(n^{ O( \Lambda ) } \cdot (\frac{ 1 }{ \eps })^{ O(\Lambda) } )$-time approach. As part of their concluding remarks, the authors  asked whether it is possible to obtain a fixed-parameter tractable PTAS, namely, one admitting an $f_1( \Lambda, \frac{ 1 }{ \eps } ) \cdot |{\cal I}|^{ f_2( 1/\eps ) }$-time implementation, where $|{\cal I}|$ stands for our input length in its binary representation. 

In Section~\ref{sec:approx_scheme_Lambda}, we resolve the above-mentioned question, in a much stronger sense: As long as $\Lambda$ is polynomial in the input size (rather than a mere constant), we prove that the optimal peak inventory level can be efficiently approached within any degree of accuracy, as formally stated in  Theorem~\ref{thm:main_result_lambda} below. For convenience, we design our LP-based approximation scheme while shooting for a success probability of $\frac{1}{2}$; the latter can be arbitrarily amplified via independent repetitions. Yet another important remark is that this result can easily be adapted to the continuous problem formulation, as well as to exponentially-large identical time intervals, via relatively simple reductions to the discrete setting (see Sections~\ref{subsec:discretize} and~\ref{subsec:reduction_cont_disc}).

\begin{theorem} \label{thm:main_result_lambda}
For any $\eps > 0$, the discrete inventory staggering problem can be approximated within factor $1 + \eps$ of optimal. The running time of our randomized algorithm is $O( \Lambda^{ \tilde{O}( 1/\eps^3 ) } \cdot  |{\cal I}|^{ O(1) } )$, and it is successful with probability at least $\frac{ 1 }{ 2 }$.
\end{theorem}

\paragraph{Main result 4: Approximation scheme in terms of distinct time intervals.} The next thread of work we consider is that of approximating inventory staggering instances characterized by constantly-many distinct time intervals, noting that this setting will play an important role within the algorithmic advances presented in future sections. In spite of their misleadingly simple structure, such instances continue to elude truly near-optimal solution methods, with non-trivial guarantees currently available only for two distinct time intervals. In this scenario, \citet{TeoOT98} showed that continuous inventory staggering can be efficiently approximated within factor $\frac{ 4 }{ 3 }$, when one time interval is an integer multiple of the other. Later on, \citet{HumST05} eliminated the latter assumption, attaining an approximation ratio of roughly $1.34312$ via a generalized Homer’s policy \citep{Homer1966}.

In Section~\ref{sec:approx_scheme_const_intervals}, we improve on the above-mentioned guarantees, further lifting these results beyond the two-interval scenario, along the lines of Theorem~\ref{thm:main_result_K} below. Specifically, letting $K$ designate the underlying number of distinct time intervals, we devise a deterministic approach for discrete inventory staggering, arguing that the latter setting can be approximated within factor $1 + \eps$ in $O( 2^{ \tilde{O}( K / \eps^2 ) } \cdot | {\cal I} |^{O(1)} )$ time. Similarly to Theorem~\ref{thm:main_result_lambda}, this result seamlessly migrates to the continuous problem formulation. A particularly interesting tool we develop en route is an exact $O ( n^{ O(n) } \cdot | {\cal I} |^{ O(1) } )$-time oracle for computing the peak inventory level of a given policy, thereby resolving the open question of peak evaluation for constantly-many items.
 
\begin{theorem} \label{thm:main_result_K}
For any $\eps > 0$, the discrete inventory staggering problem can be approximated within factor $1 + \eps$ of optimal. The running time of our algorithm is $O( 2^{ \tilde{O}( K / \eps^2 ) } \cdot | {\cal I} |^{O(1)} )$.
\end{theorem}

\paragraph{Main result 5: Approximation scheme for nested instances.} Beyond the aforementioned scenarios, prior studies have uncovered non-trivial approximation guarantees under two additional structural features. The first such finding has been established for so-called nested instances; here, when the set of items is indexed such that $T_1 \leq \cdots \leq T_n$, each time interval divides its successor. This setting, which is highly compelling due to its emergence in power-of-$2$ policies (see, e.g., \citet{Roundy85, Roundy86, JacksonMM85, MuckstadtR93}) is known to admit a polynomial-time $\frac{ 15 }{ 8 }$-approximation \citep{TeoOT98}. Our main result for nested instances, whose specifics are discussed in Section~\ref{sec:approx_scheme_nested}, resides in devising a polynomial-time approximation scheme.

\begin{theorem} \label{thm:main_result_nested}
For any $\eps > 0$, the nested inventory staggering problem can be approximated within factor $1 + \eps$ of optimal. The running time of our algorithm is $O ( 2^{ \tilde{O}(1/\eps^3) } \cdot | {\cal I} |^{ O(1) } )$.
\end{theorem}

\paragraph{Main result 6: Approximation scheme for continuous pairwise coprime instances.} The second setting for which a sub-$2$-approximation has been demonstrated to exist is centered around pairwise coprime instances of the continuous inventory staggering problem. Here, all pairs $T_i \neq T_j$ of time intervals are assumed to be relatively prime, meaning that $\gcd( T_i, T_j ) = 1$. It is worth noting that, by the Chinese Remainder Theorem, the peak inventory level of any replenishment policy for the discrete problem formulation is precisely $\sum_{i \in [n]} H_i$, rendering it completely trivial. However, when  continuous inventory staggering is concerned, this property can easily be violated, making such instances particularly difficult to deal with. As outlined in Section~\ref{sec:approx_comprime}, our final contribution comes in the form of a polynomial-time approximation scheme for pairwise coprime instances of the latter formulation, improving on its existing $1.4$-approximation due to \citet{HumST05}.

\begin{theorem} \label{thm:main_result_coprime}
For any $\eps > 0$, the continuous coprime inventory staggering problem can be approximated within factor $1 + \eps$ of optimal. The running time of our algorithm is $O( \mytower_4(O(\frac{ 1 }{ \eps}), O(\frac{ 1 }{ \eps})) \cdot | {\cal I} |^{O(1)} )$.
\end{theorem}

As a side note, to avoid convoluted running time expressions, this theorem utilizes the function $\mytower_{ \beta }(h,x)$, where $\beta$ is iteratively exponentiated $h-1$ times, finally taken to the power of $x$, i.e.,
\[ \mytower_{ \beta }(h,x) ~~=~~ \underbrace{ \beta^{{\displaystyle \beta}^{\cdot^{\cdot^{\cdot^{{\displaystyle \beta}^{ \displaystyle x }}}}}} }_{ \text{$h$ times} } \ . \]
As such, $\mytower_4(O(\frac{ 1 }{ \eps}), O(\frac{ 1 }{ \eps}))$ is a function of $\frac{ 1 }{ \eps }$ and nothing more. To our knowledge, approximation schemes characterized by power-tower running times are few and far between, mostly arising as algorithmic applications of Szemer\'edi's Regularity Lemma; see, for instance, \citet{AlonDLRY94} and \citet{DukeLR95}.

\paragraph{Hardness results.}  To better understand the nature of algorithmic outcomes one could be striving for, it is worth briefly mentioning known intractability results surrounding inventory staggering. Along these
lines, \citet{GallegoSS92} were the first to rigorously investigate how plausible it is to efficiently compute optimal shift vectors, proving that
the continuous problem formulation is strongly NP-hard. Subsequently, \citet{Hall98} studied discrete inventory staggering, establishing weak NP-hardness even when all items share a common time interval of $T=2$. Recently, \citet{HochbaumR19} showed that the latter setting becomes strongly NP-hard when $T$ could be arbitrarily large. To our knowledge, whether the discrete formulation or its continuous counterpart are APX-hard in their full generality is still a major unresolved question.

%% file: TEX-Impossibility.tex
\section{Impossibility Results} \label{sec:impossible}

In this section, we explore fundamental questions about estimating the peak inventory level of a given shift vector as well as about the plausibility of identifying non-trivial solutions via groupwise synchronization. Toward these objectives, Section~\ref{subsec:impossible_prelim} presents a number of known observations that will be helpful down the road. Then, Section~\ref{subsec:peak_eval_sample} derives an exponential lower bound on the sample complexity of peak evaluation. Finally, Section~\ref{subsec:groupwise_sync} is devoted to arguing that even $\tilde{\Omega}( \sqrt{n} )$-synchronization is insufficient to obtain non-straightforward approximation guarantees.

\subsection{Auxiliary claims} \label{subsec:impossible_prelim}

\paragraph{The average-space lower bound.} Moving forward, it will be useful to keep in mind well-known lower bounds on the peak inventory levels of optimal replenishment policies for the discrete and continuous  formulations, $\opt^{ \mydisc }( {\cal I} )$ and $\opt^{ \mycont }( {\cal I} )$. These bounds follow from a simple averaging argument, and we provide their proof for completeness. To avoid cumbersome expressions, $H_{\Sigma} = \sum_{i \in [n]} H_i$ will stand for the total ordering quantity of all items.

\begin{lemma} \label{lem:avg_space_LB}
$\opt^{ \mydisc }( {\cal I} ) \geq \frac{ 1 }{ 2 } \cdot \sum_{i \in [n]} H_i \cdot (1 + \frac{ 1 }{ T_i } )$ and $\opt^{ \mycont }( {\cal I} ) \geq  \frac{ H_{\Sigma} }{ 2 }$.
\end{lemma}
\begin{proof}
Recalling that $\Lambda = \lcm( T_1, \ldots, T_n)$ is the cycle length of any replenishment policy, suppose we draw a random point $X \sim U \{ 0, 1, \ldots, \Lambda-1 \}$. Then, for any shift vector $\tau \in \bbZ^n_+$ and for any item $i \in [n]$, it is easy to verify that $I_i( \tau_i, X) \sim U \{ \frac{ H_i }{ T_i }, \frac{ 2H_i }{ T_i }, \ldots, \frac{ T_i H_i }{ T_i } \}$. Therefore, 
\[ I_{ \max }( \tau) ~~\geq~~ \ex{ I_{ \Sigma }( \tau, X ) } ~~=~~ \sum_{i \in [n]} \ex{ I_i( \tau_i, X } ~~=~~ \frac{ 1 }{ 2 } \cdot \sum_{i \in [n]} H_i \cdot \left(1 + \frac{ 1 }{ T_i } \right) \ , \]
implying that $\opt^{ \mydisc }( {\cal I} ) \geq \frac{ 1 }{ 2 } \cdot \sum_{i \in [n]} H_i \cdot (1 + \frac{ 1 }{ T_i } )$. In the continuous setting, we repeat precisely the same argument, except for sampling $X \sim U[0, \Lambda]$. As a result, $I_i( \tau_i, X) \sim U[0, H_i]$, ending up with $\opt^{ \mycont }( {\cal I} ) \geq \frac{ 1 }{ 2 } \cdot \sum_{i \in [n]} H_i = \frac{ H_{\Sigma} }{ 2 }$.
\end{proof}

\paragraph{The near-optimality regime of random shifts.} Let $\tau^R$ be a random shift vector where, for each item $i \in [n]$, we pick $\tau^R_i \sim U \{ 0, \ldots, T_i-1 \}$, independently of any other item. In what follows, we focus our attention on the next question: What conditions guarantee that $\tau^R$ is near-optimal with high probability, or at least with some positive probability? To derive such conditions, we first establish the next claim, providing an upper bound on the probability that the random inventory level of $\tau^R$ pointwise exceeds the average-space bound by a factor greater than $1 + \eps$.

\begin{lemma} \label{thm:main_result_eps_good}
For every time point $t \in [0, \Lambda] \cap \bbZ$, 
\[ \pr{ I_{ \Sigma }( \tau^R, t) \geq \frac{ 1 + \eps }{ 2 } \cdot \sum_{i \in [n]} H_i \cdot \left( 1 + \frac{ 1 }{ T_i } \right) } ~~\leq~~ \exp \left( - \frac{ \eps^2 }{ 6 } \cdot \frac{ H_{\Sigma} }{ H_{\max} } \right) \ . \]
\end{lemma}
\begin{proof}
To derive the desired claim, for every item $i \in [n]$, let $Y_i^t = I_i( \tau^R_i, t)$ be its inventory level at time $t$ in terms of the random policy ${\cal P}_{ \tau^R }$. It is easy to see that $Y_i^t \sim U \{ \frac{ H_i }{ T_i }, \frac{ 2H_i }{ T_i }, \ldots, \frac{ T_i H_i }{ T_i } \}$, and consequently, 
\begin{eqnarray*}
&& \pr{ I_{ \Sigma }( \tau^R, t) \geq \frac{ 1 + \eps }{ 2 } \cdot \sum_{i \in [n]} H_i \cdot \left( 1 + \frac{ 1 }{ T_i } \right) } \\
&& \qquad =~~ \pr{ \sum_{i \in [n]} \frac{ Y_i^t }{ H_{\max} } \geq (1 + \eps) \cdot \ex{ \sum_{i \in [n]} \frac{ Y_i^t }{ H_{\max} } } } \\
&& \qquad \leq~~ \exp \left( - \frac{ \eps^2 }{ 6 } \cdot \frac{ H_{\Sigma} }{ H_{\max} } \right) \ .
\end{eqnarray*}
The last inequality is obtained by employing the next Chernoff-Hoeffding bound (see \citet[Thm.~1.1 and Ex.~1.1]{DubhashiP09}) with respect to $\{ \frac{ Y_i^t }{ H_{\max} } \}_{i \in [n]}$: Let $Z_1, \ldots, Z_n$ be independent $[0,1]$-bounded random variables. Then, for every $\bar{\mu} \geq \expar{ \sum_{i \in [n]} Z_i }$ and $\xi \in (0,1)$, we have
\begin{equation} \label{eqn:Chernoff-Hoeffding}
\pr{ \sum_{i \in [n]} Z_i > (1 + \xi) \cdot \bar{\mu} } ~~\leq~~ \exp \left( - \frac{ \xi^2 \bar{\mu} }{ 3 } \right) \ . 
\end{equation}
\end{proof}

The first implication of Lemma~\ref{thm:main_result_eps_good} is that, when $\Lambda < \exp ( \frac{ \eps^2 }{ 6 } \cdot \frac{ H_{\Sigma} }{ H_{\max} } )$, by taking the union bound over all times point in $[0, \Lambda) \cap \bbZ$, we have $I_{ \max }( \tau^R) \leq \frac{ 1 + \eps }{ 2 } \cdot \sum_{i \in [n]} H_i \cdot ( 1 + \frac{ 1 }{ T_i } )$ with positive probability. Consequently, $\opt^{ \mydisc}( {\cal I } ) \leq \frac{ 1 + \eps }{ 2 } \cdot \sum_{i \in [n]} H_i \cdot ( 1 + \frac{ 1 }{ T_i } )$ for any such instance ${\cal I}$, nearly matching the average-space lower bound of Lemma~\ref{lem:avg_space_LB}. The second implication of this result is that, when $\Lambda < \delta \cdot \exp ( \frac{ \eps^2 }{ 6 } \cdot \frac{ H_{\Sigma} }{ H_{\max} } )$, the peak inventory level of $\tau^R$ is within factor $1 + \eps$ of optimal, with probability at least $1 - \delta$. It is worth mentioning that bounds of similar flavor have previously been obtained by \citet{CrootH13random}; however, their precise form is somewhat cumbersome to exploit in our particular setting.

\subsection{The sample complexity of peak evaluation} \label{subsec:peak_eval_sample}

In what follows, we derive Theorem~\ref{thm:LB_random_sample}, showing that $2^{ \Omega(\eps^2 n ) }$ randomly drawn points are required in order to estimate the peak inventory level of a given shift vector within factor $\frac{ 1 }{ 2 } + \eps$ with constant probability.

\paragraph{Construction.} Let us consider a discrete inventory staggering instance ${\cal I}$ with $n \geq \frac{ 2 }{ \eps }$ items, whose ordering quantities are uniform, say $H_1 = \cdots = H_n = 1$. The time intervals $T_1, \ldots, T_n$ of these items are distinct primes, each of value at least $n$; as a result, our cycle length is $\Lambda = \prod_{i \in [n]} T_i$. Focusing on the shift vector $\tau = \vec{0}$, we wish to estimate the peak inventory level of ${\cal P}_{ \vec{0} }$, which is clearly $I_{ \max }(\vec{0}) = n$. To this end, our randomized procedure operates as follows:
\begin{itemize}
    \item We independently draw $M$ time points, $X_1, \ldots, X_M$, such that $X_m \sim U \{ 0, \ldots, \Lambda - 1 \}$.

    \item Our estimate $\tilde{I}_{ \max }^M$ for $I_{ \max }( \vec{0})$ is given by the maximum value of $I_{ \Sigma }(\vec{0}, \cdot)$ across these points, i.e., $\tilde{I}_{ \max }^M = \max_{m \in [M]} I_{ \Sigma }(\vec{0}, X_m)$.
\end{itemize}

\paragraph{Analysis.} The next claim shows that, with probability at least $1 - M e^{ - \eps^2 n / 6 }$, the maximal inventory level inferred from our sample is at most $(\frac{ 1 }{ 2 } + \eps) \cdot n$. As previously noted, $I_{ \max }(\vec{0}) = n$, meaning that the number of points required to attain a $(\frac{ 1 }{ 2 } + \eps)$-estimate with constant probability is $2^{ \Omega(\eps^2 n ) }$.

\begin{lemma}
$\prpar{ \tilde{I}_{ \max }^M \geq (\frac{ 1 }{ 2 } + \eps) \cdot n} \leq M e^{ - \eps^2 n / 6 }$. 
\end{lemma}
\begin{proof}
By taking the union bound over $X_1, \ldots, X_M$, it suffices to show that $\prpar{ I_{ \Sigma }( \vec{0}, X_m) \geq (\frac{ 1 }{ 2 } + \eps) \cdot n } \leq e^{ - \eps^2 n / 6 }$ for every $m \in [M]$. To this end, letting $I_i^{X_m} = I_i( 0, X_m)$ be the inventory level of item $i$ at time $X_m$, it is easy to see that $I_i^{X_m} \sim U \{ \frac{ 1 }{ T_i }, \frac{ 2 }{ T_i }, \ldots, \frac{ T_i }{ T_i } \}$, implying in particular that $\expar{ I_i^{X_m} } = \frac{ 1 }{ 2 }(1 + \frac{ 1 }{ T_i })$. The important observation is that $\{ I_i^{X_m} \}_{i \in [n]}$ are mutually independent. To verify this claim, given a vector $s = (s_1, \ldots, s_n) \in [T_1] \times \cdots \times [T_n]$, let $t_s$ be the unique solution modulo $\Lambda$ to the system of congruences
\[ t ~~\equiv~~ T_i - s_i \ (\mymod \  T_i) \qquad \forall \, i \in [n]  \]
Recalling that $T_1, \ldots, T_n$ are distinct primes, by the Chinese Remainder Theorem (see, e.g., \citep[Sec.~2.3]{NivenZM91}), we know that $t_s$ indeed exists, and also, that it is unique (modulo $\Lambda$). Since we are considering the shift vector $\tau = \vec{0}$, it follows that $\bigwedge_{i \in [n]} \{ I_i^{X_m} = \frac{ s_i }{ T_i } \} = \{ X_m = t_s \}$, implying that $\{ I_i^{X_m} \}_{i \in [n]}$ are mutually independent, as 
\[ \pr{ \bigwedge_{i \in [n]} \left\{ I_i^{X_m} = \frac{ s_i }{ T_i } \right\} } ~~=~~ \pr{ X_m = t_s} ~~=~~ \frac{ 1 }{ \Lambda } ~~=~~  \frac{ 1 }{ \prod_{i \in [n]} T_i } ~~=~~ \prod_{i \in [n]} \pr{ I_i^{X_m} = \frac{ s_i }{ T_i } } \ . \]
Based on these observations, we derive the desired upper bound on $\prpar{ I_{ \Sigma }( \vec{0}, X_m) \geq (\frac{ 1 }{ 2 } + \eps) \cdot n }$ by noting that
\begin{eqnarray}
\pr{ I_{ \Sigma }( \vec{0}, X_m) \geq \left( \frac{ 1 }{ 2 } + \eps \right) \cdot n } & \leq & \pr{ \sum_{i \in [n]} I_i^{X_m} \geq (1 + \eps) \cdot \ex{ \sum_{i \in [n]} I_i^{X_m} } } \label{eqn:samp_comp_eq1} \\
& \leq & \exp \left( - \frac{ \eps^2 }{ 3 } \cdot \ex{ \sum_{i \in [n]} I_i^{X_m} } \right) \label{eqn:samp_comp_eq2} \\
& \leq & e^{ - \eps^2 n / 6 } \ . \label{eqn:samp_comp_eq3}
\end{eqnarray}
Here, inequality~\eqref{eqn:samp_comp_eq1} holds since $(1 + \eps) \cdot \expar{ \sum_{i \in [n]} I_i^{X_m} } = \frac{ 1 + \eps }{ 2 } \cdot \sum_{i \in [n]} (1 + \frac{ 1 }{ T_i }) \leq (\frac{1}{2} +\eps) \cdot n$, as $T_i \geq n \geq \frac{ 2 }{ \eps }$. Inequality~\eqref{eqn:samp_comp_eq2} follows by specializing the Chernoff-Hoeffding bound~\eqref{eqn:Chernoff-Hoeffding} with respect to the random variables $\{ I_i^{X_m} \}_{i \in [n]}$, which are  independent and $[0,1]$-bounded. Finally, inequality~\eqref{eqn:samp_comp_eq3} follows by recalling that $\expar{ I_i^{X_m} } = \frac{ 1 }{ 2 }(1 + \frac{ 1 }{ T_i }) \geq \frac{ 1 }{ 2 }$. 
\end{proof}

\subsection{Infeasibility of groupwise synchronization} \label{subsec:groupwise_sync}

Our second impossibility result is motivated by the recent work of \citet{Segev24} on economic warehouse lot scheduling, where among other ideas, pairwise synchronization has been crucial for improving on the  long-standing $2$-approximation in this context \citep{Anily91, GallegoQS96}. Migrating this notion to inventory staggering, one would operate along the following lines: 
\begin{itemize}
    \item First, for each pair of items $i \neq j$, we compute an optimal shift vector $\tau^{ij}$. This step is implementable in polynomial time, given the existence of closed-form solutions for two-item instances \citep{HartleyT82, MurthyBR03}. Our hope is to discover that, for a significant portion of these pairs, a peak inventory level of $I_{\max}( \tau^{ij} ) \leq (1-\delta_{ij}) \cdot (H_i + H_j)$ can be attained, for some $\delta_{ij} > 0$.

    \item Then, the big unknown is whether there exists a partition ${\cal M}$ of the overall set of items into pairs, such that $\sum_{(i,j) \in {\cal M}} I_{\max}( \tau^{ij} ) \leq (1- \delta) \cdot H_{\Sigma}$, for some absolute constant $\delta > 0$. Combined with Lemma~\ref{lem:avg_space_LB}, such a result would immediately lead to a $2(1-\delta)$-approximation for the discrete inventory staggering problem in its full generality.
\end{itemize}

Unfortunately, we argue that this approach cannot succeed for pairs of items, for triplets, or for subsets of much larger size. As formally stated in Theorem~\ref{thm:LB_group_sync}, the remainder of this section is dedicated to presenting an inventory staggering instance ${\cal I}$ where, on the one hand, $\opt^{ \mydisc }( {\cal I} ) \leq (1 + 2\eps) \cdot \frac{ H_{\Sigma} }{ 2 }$, and on the other hand, $\opt^{ \mydisc }( {\cal I}_{\hat{\cal F}} ) \geq (1 - \eps) \cdot H( \hat{\cal F} )$ for every subset $\hat{\cal F}$ of $O( \frac{ \sqrt{n} }{ \log n } )$ items. Consequently, even when one is capable of computing optimal shift vectors for subsets of nearly $\sqrt{n}$ items, simply gluing them together across all subsets would still be far from the global optimum by a factor of $2-O(\eps)$.

\paragraph{Construction.} To describe our construction, it will be convenient to utilize a well-known result regarding sparse combinatorial designs. Specifically, suppose that ${\cal F}$ is a family of $q$-element subsets of $[n]$. For $r \leq q$, we say that ${\cal F}$ is $r$-sparse when any two subsets in ${\cal F}$ intersect in at most $r-1$ elements; ${\cal M}(n,q,r)$ will designate the maximum size of such a family. The next result  is taken from the work of \citet[Sec.~2.2]{AlonMMV12}, noting that bounds in this spirit have also been suggested earlier, for example, by \citet{HartmanMS82} and by \citet{NisanW94}.

\begin{theorem} \label{thm:AlonMMV12}
${\cal M}( q^2, q, r ) \geq q^r$, for every prime power $q$ and $2 \leq r \leq q$.
\end{theorem}

Now, given an error parameter $\eps > 0$, the inventory staggering instance ${\cal I}$ we consider is constructed as follows:
\begin{itemize}
    \item For a prime $q \geq e^{ 1/(2\eps^2) }$, let $K = q^2$, and let $p_1, \ldots, p_K$ be a collection of distinct primes within $[K,2K \ln K]$. By the Prime Number Theorem (see, e.g., \citet[Sec.~7]{DavenportM13}), for sufficiently large $K$, we indeed have at least $K$ such primes, with room to spare.

    \item {\em Items}: By Theorem~\ref{thm:AlonMMV12}, we know that ${\cal M}(K, \sqrt{K}, 3) \geq K^{3/2}$, implying the weaker claim that  there exists a $3$-sparse family ${\cal F}$ of $\sqrt{K}$-element subsets of $[K]$, with $|{\cal F}| = \lceil 12K \ln^2 K \rceil$. For every subset $S \in {\cal F}$, we introduce a corresponding item $i_S$, meaning that $n = | {\cal F} |$.

    \item {\em Time intervals and ordering quantities}: For every item $i_S$, the time interval between successive $i_S$-orders is $T_{i_S} = \prod_{k \in S} p_k$, with an ordering quantity of $H_{i_S} = 1$.
\end{itemize}

\paragraph{Upper bound on global optimum.} With respect to this instance, we begin by arguing that its optimal peak inventory level is essentially $\frac{ n }{ 2 }$, up to $\eps$-dependent terms. This finding will be deduced by showing that ${\cal I}$ meets our sufficient condition for nearly matching the average-space lower bound of Lemma~\ref{lem:avg_space_LB}. 

\begin{lemma} \label{lem:UB_groupwise}
$\opt^{ \mydisc }( {\cal I} ) \leq (1 + 2\eps) \cdot \frac{ n }{ 2 }$.
\end{lemma}
\begin{proof}
Based on the concluding paragraph of Section~\ref{subsec:impossible_prelim}, when $\Lambda < \exp ( \frac{ \eps^2 }{ 6 } \cdot \frac{ H_{\Sigma} }{ H_{\max} } )$, we have 
\[ \opt^{ \mydisc }( {\cal I} ) ~~\leq~~ \frac{ 1 + \eps }{ 2 } \cdot \sum_{S \in {\cal F}} H_{i_S} \cdot \left( 1 + \frac{ 1 }{ T_{i_S} } \right) ~~\leq~~ \frac{ (1 + \eps) ( 1 + e^{ -2/\eps } ) }{ 2 } \cdot \sum_{S \in {\cal F}}  H_{i_S} ~~\leq~~ (1 + 2\eps) \cdot \frac{ n }{ 2 } \ . \]
To verify the above-mentioned condition for sufficiently large $K$, note that since $p_1, \ldots, p_K$ are distinct primes within $[K,2K \ln K]$, we have
\begin{eqnarray}
\Lambda & = & \lcm \left\{ T_{i_S} : S \in {\cal F} \right\} \nonumber \\
& \leq & \prod_{k \in [K]} p_k \nonumber \\
& < & (2K \ln K)^{ K } \nonumber \\
& \leq & e^{ 2K \ln K } \nonumber \\
& \leq & \exp \left( \frac{ \eps^2 n }{ 6 } \right) \label{eq:proof_lem_UB_groupwise_1} \\
& = & \exp \left( \frac{ \eps^2 }{ 6 } \cdot \frac{ H_{\Sigma} }{ H_{\max} } \right) \ . \label{eq:proof_lem_UB_groupwise_2}
\end{eqnarray}
Here, inequality~\eqref{eq:proof_lem_UB_groupwise_1} holds since $K \geq e^{ 1/\eps^2 }$ and $n = \lceil 12K \ln^2 K \rceil$, whereas equality~\eqref{eq:proof_lem_UB_groupwise_2} is obtained by noting that $H_{\Sigma} = n$ and $H_{\max} = 1$, since $H_{i_S} = 1$ for every $S \in {\cal F}$.
\end{proof}

\paragraph{Lower bound on $\bs{ O( \frac{ \sqrt{n} }{ \log n } )}$-wise optimum.} Now, let us consider any subset $\hat{\cal F} \subseteq {\cal F}$ of size at most $\frac{ \sqrt{K} }{ 2 }$, noting that $\sqrt{K} = \Theta(  \frac{ \sqrt{n} }{ \log n } )$. We make use of ${\cal I}_{\hat{\cal F}}$ to designate the instance obtained by restricting ${\cal I}$ to this subset. To conclude the proof of Theorem~\ref{thm:LB_group_sync}, we show that with respect to this instance, the optimal peak inventory level is nearly $| \hat{\cal F} |$, which identifies with the combined ordering quantity across all $\hat{\cal F}$-items, and therefore, attainable by any shift vector.

\begin{lemma} \label{lem:LB_subsets_F}
$\opt^{ \mydisc }( {\cal I}_{\hat{\cal F}} ) \geq (1 - \eps) \cdot | \hat{\cal F} |$.
\end{lemma}

To establish this bound, we begin by deriving an auxiliary number-theoretic claim. Here, the primes $p_1, \ldots, p_L$ are said to be unique divisors of the positive integers $n_1, \ldots, n_L$ when, for every $\ell \in [L]$, the next two properties are satisfied: (1)~$p_{\ell}$ is a divisor of $n_{\ell}$; and (2)~$p_{\ell}$ is a not a divisor of $n_k$, for every $k \neq \ell$. 

\begin{lemma} \label{lem:small_remainders}
Suppose that $p_1, \ldots, p_L$ are unique divisors of $n_1, \ldots, n_L$. Then, for any $\tau \in \bbZ^L$, there exists an integer $t$ such that, for every $\ell \in [L]$,
\[ (t - \tau_{\ell}) \ \mymod \ n_{\ell} ~~\in~~ \left[ 0, \frac{ n_{\ell} }{ p_{\ell} } - 1 \right] \ . \]
\end{lemma}
\begin{proof}
In what follows, we make use of $\alpha_{\ell} \geq 1$ to denote the largest integer for which $p_{\ell}^{ \alpha_{\ell} }$ is a divisor of $n_{\ell}$. With this notation, we will actually prove a stronger result, showing that there exists an integer $t$ such that $(t - \tau_{\ell}) \ \mymod \ n_{\ell} \in [ 0, \frac{ n_{\ell} }{ p_{\ell}^{ \alpha_{\ell} } } - 1 ]$, for every $\ell \in [L]$.

To this end, for every $\ell \in [L]$ and $0 \leq r \leq p_{\ell}^{ \alpha_{\ell} }-1$, consider the system of congruences
\[ (F_{\ell,r}) \left\{ \begin{array}{l}
t \equiv 0 \ (\mymod \ \frac{ n_{\ell} }{ p_{\ell}^{ \alpha_{\ell} } }) \\
t \equiv r \ (\mymod \ p_{\ell}^{ \alpha_{\ell} })  
\end{array} \right. \]
Since $\frac{ n_{\ell} }{ p_{\ell}^{ \alpha_{\ell} } }$ and $p_{\ell}^{ \alpha_{\ell} }$ are relatively prime, by the Chinese Remainder Theorem, $(F_{\ell,r})$ has a unique integer solution $t_{\ell,r}$ modulo $n_{\ell}$. It is easy to verify that $t_{\ell,0}, \ldots, t_{\ell,p_{\ell}^{ \alpha_{\ell} }-1}$ are distinct multiples of $\frac{ n_{\ell} }{ p_{\ell}^{ \alpha_{\ell} } }$ in $[0, n_{\ell}-1]$, and clearly, there are exactly $p_{\ell}^{ \alpha_{\ell} }$ such multiples. Therefore, $(t_{\ell,0} - \tau_{\ell}) \ \mymod \ n_{\ell}, \ldots, (t_{\ell,p_{\ell}^{ \alpha_{\ell} }-1} - \tau_{\ell}) \ \mymod \ n_{\ell}$ necessarily stab each segment of $\frac{ n_{\ell} }{ p_{\ell}^{ \alpha_{\ell} } }$ consecutive integers in $\bbZ_{ n_{\ell} }$. In particular, there exists some index $r_{\ell}$ for which $(t_{\ell,r_{\ell}} - \tau_{\ell}) \ \mymod \ n_{\ell} \in [ 0, \frac{ n_{\ell} }{ p_{\ell}^{ \alpha_{\ell} } } - 1 ]$.

Now, to identify the desired integer $t$, letting $M = \frac{ \lcm( n_1, \ldots, n_L ) }{ \prod_{\ell \in [L]} p_{\ell}^{ \alpha_{\ell} } }$, consider the system of congruences
\[ \left\{ \begin{array}{ll}
t \equiv 0 \ (\mymod \ M) \\
t \equiv r_{\ell} \ (\mymod \ p_{\ell}^{ \alpha_{\ell} }) \quad & \forall \, \ell \in [L]
\end{array} \right. \]
Once again, since $M, p_{1}^{ \alpha_{1} }, \ldots, p_{L}^{ \alpha_{L} }$ are pairwise coprime, this system has a unique integer solution $t$ modulo $\lcm( n_1, \ldots, n_L )$. We conclude the proof by observing that $(t - \tau_{\ell}) \ \mymod \ n_{\ell} \in [ 0, \frac{ n_{\ell} }{ p_{\ell}^{ \alpha_{\ell} } } - 1 ]$ for every $\ell \in [L]$. Indeed,  since $\frac{ n_{\ell} }{ p_{\ell}^{ \alpha_{\ell} } }$ divides $M$, we know in particular that $t$ is a solution to $(F_{\ell,r_{\ell}})$. However, as explained above, the latter system has a unique solution $t_{\ell,r_{\ell}}$ modulo $n_{\ell}$, implying that by our choice of $r_{\ell}$,
\[ (t - \tau_{\ell}) \ \mymod \ n_{\ell} ~~\equiv~~ (t_{\ell,r_{\ell}} - \tau_{\ell}) \ \mymod \ n_{\ell} ~~\in~~ \left[ 0, \frac{ n_{\ell} }{ p_{\ell}^{ \alpha_{\ell} } } - 1 \right] \ . \]
\end{proof}

Given this claim, we proceed by arguing that $\opt^{ \mydisc }( {\cal I}_{\hat{\cal F}} ) \geq (1 - \eps) \cdot | \hat{\cal F} |$. For this purpose, let $\tau^*$ be an optimal shift vector with respect to ${\cal I}_{\hat{\cal F}} $, meaning that $I_{\max}( \tau^* ) = \opt^{ \mydisc }( {\cal I}_{\hat{\cal F}}  )$. Focusing on a single subset $S \in \hat{\cal F}$, we observe  that there are at least two elements in $S$ that do not appear in any other subset in $\hat{\cal F}$, since
\[ \left| S \setminus \left(  \bigcup_{\hat{S} \in \hat{\cal F} \setminus \{ S \} } \hat{S} \right) \right| ~~\geq~~ |S| - \sum_{\hat{S} \in \hat{\cal F} \setminus \{ S \} } \left| S \cap \hat{S} \right| ~~\geq~~ \sqrt{K} - 2 \cdot \left( \frac{ \sqrt{K} }{ 2 }-1 \right) ~~=~~ 2 \ ,  \]
where the second inequality holds since ${\cal F}$ is a $3$-sparse family of $\sqrt{K}$-element subsets and since $| \hat{\cal F} | \leq \frac{ \sqrt{K} }{ 2 }$. We make use of $k_S \in S$ to denote one of these two elements.

As a result of this observation,  it follows that the primes $(p_{ k_S })_{ S \in \hat{\cal F} }$ are unique divisors of $(T_{ i_S })_{ S \in \hat{\cal F} }$. Therefore, by Lemma~\ref{lem:small_remainders}, there exists an integer $t$ such that $(t - \tau^*_{i_S}) \ \mymod \ T_{i_S} \in [ 0, \frac{ T_{i_S} }{ p_{ k_S } } - 1 ]$, for every $S \in \hat{\cal F}$. In other words, with respect the policy ${\cal P}_{\tau^*}$, the time elapsing between the last $i_{S}$-order in $[0,t]$ and the point $t$ itself is at most $\frac{ T_{i_S} }{ p_{ k_S } }$. In turn, the inventory level of this item at time $t$ is
\[ I_{i_S}( \tau^*_{ i_S } , t) ~~\geq ~~ H_{i_{S}} \cdot \left( 1 - \frac{ T_{i_S} / p_{ k_S } }{ T_{i_{S}} } \right) ~~=~~ 1 - \frac{ 1 }{ p_{ k_S } } ~~\geq~~ 1 - \eps \ , \]
where the last inequality holds since $p_{ k_S } \geq K \geq e^{ 1/\eps^2 }$. In conclusion, 
\[ \opt^{ \mydisc }( {\cal I}_{\hat{\cal F}} ) ~~=~~ I_{\max}( \tau^* ) ~~\geq~~ I_{\Sigma}( \tau^* , t) ~~=~~ \sum_{S \in \hat{\cal F}} I_{ i_S }( {\tau^*_{i_{S}}}, t) ~~\geq~~ (1 - \eps) \cdot | \hat{\cal F} | \ . \]

%% file: TEX-Approx-Cycle-Length.tex
\section{Approximation Scheme in Terms of Cycle Length} \label{sec:approx_scheme_Lambda}

This section is dedicated to presenting a randomized algorithm for discrete inventory staggering, showing that as long as our cycle length $\Lambda = \lcm( T_1, \ldots, T_n)$ is not exponentially large, the optimal peak inventory level can be efficiently approached within any degree of accuracy. Formally, as  stated in  Theorem~\ref{thm:main_result_lambda}, we will argue that for any $\eps \in (0, 1)$, the discrete setting can be approximated within factor $1 + \eps$ of optimal via an $O( \Lambda^{ \tilde{O}( 1/\eps^3 ) } \cdot  |{\cal I}|^{ O(1) } )$-time  algorithm, which is successful with probability at least $\frac{ 1 }{ 2 }$. 

\subsection{\texorpdfstring{$\bs{O( \frac{ 1 }{ \eps } )}$}{}-sized discretization sets} \label{subsec:discretize}

For this purpose, rather than allowing the shift $\tau_i$ of each item $i \in [n]$ to take arbitrary integer values in $[0, T_i]$, we will restrict our choice to an $O( \frac{ 1 }{ \eps } )$-sized set of values ${\cal D}_i$ as follows.

\paragraph{Construction of $\bs{{\cal D}_i}$.} Given an error parameter $\eps \in (0,1)$, we say that item $i$ is large when $T_i > \frac{ 1 }{ \eps }$; in the opposite case, this item is said to be small. From this point on, ${\cal L}$ and ${\cal S}$ will respectively denote the sets of large and small items. We proceed by associating each item $i \in [n]$ with a discrete set of values ${\cal D}_i$, defined according to the next case disjunction:
\begin{itemize}
    \item {\em When $i \in {\cal S}$}: Here, we keep our set of options as ${\cal D}_i = \{ 0, 1, \ldots, T_i \}$.

    \item {\em When $i \in {\cal S}$}: In this scenario, on top of the singular value $T_i$, only rounded-down multiples of $\eps T_i$ will be allowed, specifically meaning that  ${\cal D}_i = \{  \lfloor \eps T_i \rfloor, \lfloor 2\eps T_i \rfloor, \ldots, \lfloor \lfloor \frac{ 1 }{ \eps } \rfloor \cdot \eps T_i \rfloor , T_i \}$.    
\end{itemize}
In any case, we clearly have $| {\cal D}_i | \leq \lfloor \frac{ 1 }{ \eps } \rfloor + 1$. With these definitions, a shift vector $\tau = (\tau_1, \ldots, \tau_n)$ is called ${\cal D}$-restricted when $\tau_i \in {\cal D}_i$ for every item $i$.

\paragraph{Analysis.} For the remainder of this section, $\tau^*$ will stand for some fixed optimal shift vector; without loss of generality, we assume that $\tau_i^* \in [0,T_i-1] \cap \bbZ$ for every item $i \in [n]$. The next claim shows that, in terms of their peak inventory level, ${\cal D}$-restricted shift vectors are capable of competing against the unrestricted optimal vector $\tau^*$.

\begin{lemma} \label{lem:D_restricted_good}
There exists a ${\cal D}$-restricted shift vector $\tau$ for which $I_{ \max }(\tau) \leq (1 + 4\eps) \cdot I_{ \max }(\tau^*)$.
\end{lemma}
\begin{proof}
Let us consider the shift vector $\tau$, defined by setting $\tau_i = \lceil \tau_i^* \rceil^{ {\cal D}_i }$ for every item $i \in [n]$. Here, $\lceil \cdot \rceil^{ {\cal D}_i }$ is an operator that rounds its argument up to the nearest point in ${\cal D}_i$, implying that $\tau$ is ${\cal D}$-restricted. With respect to this definition, we clearly have $\tau_i = \tau_i^*$ for every item $i \in {\cal S}$, meaning in particular that $I_i(\tau_i,t ) = I_i( \tau_i^*,t )$ at any time point $t \in [0, \Lambda]$. For every item $i \in {\cal L}$, while we generally have $\tau_i \geq \tau_i^*$, note that
\begin{eqnarray*}
I_i({\tau_i},t ) & \leq & I_i({\tau_i^*},t ) + \frac{ H_i }{ T_i } \cdot ( \lceil \tau_i^* \rceil^{ {\cal D}_i } - \tau_i^*) \\
& \leq & I_i({\tau_i^*},t ) + \frac{ H_i }{ T_i } \cdot ( \eps T_i + 1) \\
& \leq & I_i({\tau_i^*},t ) + 2\eps H_i \ ,
\end{eqnarray*}
where the last inequality holds since $T_i > \frac{ 1 }{ \eps }$. By summing these relations over all items, we get $I_{ \Sigma }( \tau,t) \leq I_{ \Sigma }( \tau^*,t) + 2\eps H_{ \Sigma }$, and therefore, $I_{ \max }( \tau) \leq I_{ \max }( \tau^*) + 2\eps H_{ \Sigma } \leq (1 + 4\eps) \cdot I_{ \max }( \tau^*)$, noting that the last inequality follows from Lemma~\ref{lem:avg_space_LB}. 
\end{proof}

\subsection{The LP-rounding algorithm} \label{subsec:alg_Lambda}

\paragraph{Step 1: Guessing the peak inventory level.} Let $\tau^{\cal D}$ be a ${\cal D}$-restricted near-optimal shift vector, whose existence is guaranteed by Lemma~\ref{lem:D_restricted_good}, meaning that $I_{ \max }( \tau^{\cal D} ) \leq (1 + 4\eps) \cdot I_{ \max }( \tau^*)$. As our first algorithm step, we guess an over-estimate $\topt$ for the peak inventory level $I_{ \max }( \tau^{\cal D})$, such that $\topt \in [1, 1+\eps] \cdot I_{ \max }( \tau^{\cal D})$. Since we clearly have $I_{ \max }( \tau^{\cal D}) \leq H_{\Sigma}$, and since $I_{ \max }( \tau^{\cal D}) \geq I_{ \max }( \tau^*) \geq \frac{ H_{\Sigma} }{ 2 }$ by Lemma~\ref{lem:avg_space_LB}, there are only $O( \frac{ 1 }{ \eps } )$ candidate values to play the role of $\topt$.

\paragraph{Step 2: Guessing heavy contributions.} Letting $\delta = \frac{ \eps^3 }{ 36 \ln (2\Lambda) }$, we say that item $i \in [n]$ is heavy when $H_i \geq \delta H_{\Sigma}$; otherwise, this item is said to be light. In what follows, ${\cal H}$ and ${\cal L}$ will respectively denote the sets of heavy and light items, noting that $| {\cal H} | \leq \frac{ 1 }{ \delta }$. Our next step consists of guessing the shift $\tau^{\cal D}_i \in {\cal D}_i$ of every item $i \in {\cal H}$ with respect to $\tau^{\cal D}$. Over all heavy items, the number of required guesses is only 
\[ \prod_{i \in {\cal H}} |{\cal D}_i| ~~\leq~~ \left( \left\lfloor \frac{ 1 }{ \eps } \right\rfloor + 1 \right)^{ |{\cal H}| } ~~=~~ \left( \frac{ 1 }{ \eps } \right)^{ O(1/\delta) } ~~=~~ \Lambda^{ \tilde{O}( 1/\eps^3 ) } \ . \] 

\paragraph{Step 3: The feasibility LP.} With these parameters in place, we proceed by considering the following linear feasibility problem:
\begin{equation} \label{eqn:feasibility_LP} \tag{LP}
\begin{array}{lll}
(1) & {\displaystyle \sum_{i \in [n]} \sum_{\tau \in {\cal D}_i} I_{i \tau}^t x_{i\tau} \leq \topt} \qquad  & \forall \, t \in [0, \Lambda] \cap \bbZ \\
(2) & {\displaystyle \sum_{\tau \in {\cal D}_i} x_{i \tau} = 1} & \forall \, i \in [n] \\
(3) & x_{i \tau_i^{\cal D}} = 1 & \forall \, i \in {\cal H} \\
(4) & x_{i \tau} \geq 0 & \forall \, i \in [n], \, \tau \in {\cal D}_i
\end{array}    
\end{equation}
In an integer-valued solution to this formulation, each decision variable $x_{i\tau}$ is necessarily binary, indicating whether we choose a shift of $\tau \in {\cal D}_i$ for item $i$. As such, constraint~(1) states that the combined inventory level at any point $t$ is at most $\topt$. Here, we make use of $I_{i \tau}^t$ as a shorthand notation for $I_i( \tau, t)$, which is the inventory level of item $i$ at time $t$ with respect to the policy ${\cal P}^i_{\tau}$. Constraint~(2) can be understood as choosing exactly one of the shifts in ${\cal D}_i$ for each item $i$. Finally, constraint~(3) forces any feasible solution to be aligned with our guesses for the shifts of heavy items. It is not difficult to verify that~\eqref{eqn:feasibility_LP} is  feasible; for example, by setting $x_{i \tau} = 1$ if and only if $\tau_i^{\cal D} = \tau$, for every item $i \in [n]$ and shift $\tau \in {\cal D}_i$, we obtain a feasible solution.

\paragraph{Step 4: Randomized rounding.} Let $x^*$ be a feasible solution to~\eqref{eqn:feasibility_LP}. For each item $i \in [n]$, we draw its  random shift $\tau^R_i$, such that each $\tau  \in {\cal D}_i$ is selected with probability $x_{i \tau}^*$. This choice is made independently of any other item. As a side note, due to constraints~(2) and~(4), the quantities $\{ x_{i \tau}^* \}_{\tau \in {\cal D}_i}$ are all non-negative and sum up to $1$, meaning that they can indeed be utilized as probabilities.

\paragraph{Step 5: Peak evaluation.} Due to the guessing procedure described in steps~1 and~2, we have constructed up until now a random collection of $\Lambda^{ \tilde{O}( 1/\eps^3 ) }$ shift vectors. Out of these vectors, we return one whose peak inventory level is minimal. Even though the question of efficient peak evaluation is still wide open (see Section~\ref{subsec:prev_new_results}), it can naively be resolved in $\Lambda n^{ O(1) }$ time, simply by enumerating over all integer time points in $[0, \Lambda]$. 

\subsection{Analysis}

As stated in Lemma~\ref{lem:UB_single_time} below, we proceed by proving that, when steps~3-5 are employed with the correct set of guesses, the inventory level of our random policy ${\cal P}_{ \tau^R }$ at any given time point is at most $(1 + 2\eps) \cdot \topt$, with probability at least $1 - \frac{ 1 }{ 2\Lambda }$. By taking the union bound over all time points $t \in [0, \Lambda-1] \cap \bbZ$, we infer that $\prpar{ I_{\max}( \tau^R) \leq (1 + 2\eps) \cdot \topt } \geq \frac{ 1 }{ 2 }$. As explained in step~1, we know that $\topt \leq (1+\eps) \cdot I_{ \max }( \tau^{\cal D} )$ and that $I_{ \max }( \tau^{\cal D}) \leq (1 + 4\eps) \cdot I_{ \max }( \tau^*)$. Therefore, with probability at least $\frac{ 1 }{ 2 }$, the peak inventory level of the random policy ${\cal P}_{ \tau^R }$ is within factor $1 + 29\eps$ of optimal. 

\begin{lemma} \label{lem:UB_single_time}
For every time point $t \in [0, \Lambda] \cap \bbZ$, 
\[ \pr{ I_{ \Sigma }( \tau^R, t) \geq (1 + 2\eps) \cdot \topt } ~~\leq~~ \frac{ 1 }{ 2\Lambda } \ . \]
\end{lemma}
\begin{proof}
For every item $i \in [n]$ and shift $\tau \in {\cal D}_i$, we introduce the random variable $X_{i\tau}$, as an indicator of the event $\{ \tau_i^R = \tau \}$. In addition, the random variable $Y_i^t$ will be defined as $Y_i^t = \sum_{\tau \in {\cal D}_i} I_{i \tau}^t X_{i\tau}$, which is the inventory level of item $i$ at time $t$ in terms of the random policy ${\cal P}_{ \tau^R }$. As such, the combined inventory level of ${\cal P}_{ \tau^R }$ at time $t$ can be written as $I_{ \Sigma }( \tau^R, t) = \sum_{i \in [n]} Y_i^t$. With this notation, the bound we wish to prove is equivalent to 
\begin{equation} \label{eqn:lem_UB_single_time_equiv}
\pr{ \sum_{i \in [n]} Y_i^t \geq (1 + 2\eps) \cdot \topt } ~~\leq~~ \frac{ 1 }{ 2\Lambda } \ .  
\end{equation}
To argue about $\sum_{i \in [n]} Y_i^t$, let us decompose this summation into the contributions of heavy and light items, i.e., $\sum_{i \in [n]} Y_i^t = \sum_{i \in {\cal H}} Y_i^t + \sum_{i \in {\cal L}} Y_i^t$. The next two claims, whose proofs are provided in  Section~\ref{subsec:proof_clm_bound_heavy_light}, separately treat each of these terms.

\begin{claim} \label{clm:bound_heavy}
$\sum_{i \in {\cal H}} Y_i^t = \sum_{i \in {\cal H}} \sum_{\tau \in {\cal D}_i} I_{i \tau}^t x_{i\tau}^*$ almost surely.  
\end{claim}

\begin{claim} \label{clm:bound_light}
$\prpar{ \sum_{i \in {\cal L}} Y_i^t \geq \max \{ \eps \topt, (1 + \eps) \cdot \sum_{i \in {\cal L}} \sum_{\tau \in {\cal D}_i} I_{i \tau}^t x_{i\tau}^* \} } \leq \frac{ 1 }{ 2\Lambda}$.  
\end{claim}

With these claims in place, we proceed to prove~\eqref{eqn:lem_UB_single_time_equiv} by observing that
\begin{eqnarray}
&& \pr{ \sum_{i \in [n]} Y_i^t \geq (1 + 2\eps) \cdot \topt } \nonumber \\
&& \qquad =~~ \pr{ \sum_{i \in {\cal H}} Y_i^t + \sum_{i \in {\cal L}} Y_i^t \geq (1 + 2\eps) \cdot \topt } \nonumber \\
&& \qquad =~~ \pr{ \sum_{i \in {\cal L}} Y_i^t \geq (1 + 2\eps) \cdot \topt - \sum_{i \in {\cal H}} \sum_{\tau \in {\cal D}_i} I_{i \tau}^t x_{i\tau}^* } \label{eqn:lem_UB_single_time_1} \\
&& \qquad \leq~~ \pr{ \sum_{i \in {\cal L}} Y_i^t \geq \eps \topt + (1 + \eps) \cdot\sum_{i \in {\cal L}} \sum_{\tau \in {\cal D}_i} I_{i \tau}^t x_{i\tau}^* } \label{eqn:lem_UB_single_time_2} \\
& & \qquad \leq~~ \pr{ \sum_{i \in {\cal L}} Y_i^t \geq \max \left\{ \eps \topt, (1 + \eps) \cdot \sum_{i \in {\cal L}} \sum_{\tau \in {\cal D}_i} I_{i \tau}^t x_{i\tau}^* \right\} } \nonumber \\
& & \qquad \leq~~ \frac{ 1 }{ 2\Lambda} \ . \label{eqn:lem_UB_single_time_3} 
\end{eqnarray}
Here, equality~\eqref{eqn:lem_UB_single_time_1} follows from Claim~\ref{clm:bound_heavy}. Inequality~\eqref{eqn:lem_UB_single_time_2} holds since $\sum_{i \in {\cal H}} \sum_{\tau \in {\cal D}_i} I_{i \tau}^t x_{i\tau}^* + \sum_{i \in {\cal L}} \sum_{\tau \in {\cal D}_i} I_{i \tau}^t x_{i\tau}^* \leq \topt$, by constraint~(1) of~\eqref{eqn:feasibility_LP}. Finally, inequality~\eqref{eqn:lem_UB_single_time_3} is precisely the result stated in Claim~\ref{clm:bound_light}.
\end{proof}

\subsection{Additional proofs} \label{subsec:proof_clm_bound_heavy_light}

\paragraph{Proof of Claim~\ref{clm:bound_heavy}.} The important observation is that, for every item $i \in {\cal H}$, we have $Y_i^t = \sum_{\tau \in {\cal D}_i} I_{i \tau}^t X_{i\tau} = \sum_{\tau \in {\cal D}_i} I_{i \tau}^t x_{i\tau}^*$. Here, the second equality is obtained by noting that $X_{i \tau} = 1 [ \tau^R_i = \tau ] = x_{i\tau}^*$ for every $\tau \in {\cal D}_i$, since constraint~(3) is forcing us to choose $\tau^R_i = \tau^{\cal D}_i$ with probability $x^*_{i \tau^{\cal D}_i} = 1$. It follows that $\sum_{i \in {\cal H}} Y_i^t = \sum_{i \in {\cal H}} \sum_{\tau \in {\cal D}_i} I_{i \tau}^t x_{i\tau}^*$ almost surely.

\paragraph{Proof of Claim~\ref{clm:bound_light}.} Our proof considers two cases, depending on the relation between $\expar{ \sum_{i \in {\cal L}} Y_i^t }$ and $\topt$.

\paragraph{Case 1: $\bs{\expar{ \sum_{i \in {\cal L}} Y_i^t } \leq \frac{2 \eps}{ 3 } \cdot \topt}$.} In this scenario,
\begin{eqnarray}
\pr{ \sum_{i \in {\cal L}} Y_i^t \geq \eps \topt } & = & \pr{ \sum_{i \in {\cal L}} \frac{ Y_i^t }{ \delta H_{\Sigma} } \geq \left( 1 + \frac{ 1 }{ 2 } \right) \cdot \frac{ \frac{2 \eps}{ 3 } \cdot \topt }{ \delta H_{\Sigma}} } \nonumber \\
& \leq & \exp \left( - \frac{ \eps }{ 18 \delta} \cdot \frac{ \topt }{ H_{\Sigma}}  \right) \label{eqn:Nproof_clm_bound_light_1} \\
& \leq & \exp \left( - \frac{ \eps }{ 36\delta }  \right) \label{eqn:Nproof_clm_bound_light_2} \\
& \leq & \frac{ 1 }{ 2 \Lambda } \ . \label{eqn:Nproof_clm_bound_light_3} 
\end{eqnarray}
Here, inequality~\eqref{eqn:Nproof_clm_bound_light_1} follows by specializing the Chernoff-Hoeffding bound~\eqref{eqn:Chernoff-Hoeffding} with respect to the random variables $\{ \frac{ Y_i^t }{ \delta H_{\Sigma} } \}_{i \in {\cal L}}$, which are independent and $[0,1]$-bounded. Indeed, the former property is satisfied since the random shift of each item is independently selected, whereas the latter property is obtained by observing that $Y_i^t \leq H_i \leq \delta H_{\Sigma}$, since item $i$ is light. In addition, by the case hypothesis, $\expar{ \sum_{i \in {\cal L}} Y_i^t } \leq \frac{2 \eps}{ 3 } \cdot \topt$. Inequality~\eqref{eqn:Nproof_clm_bound_light_2} is obtained by noting that $\topt \geq I_{ \max }( \tau^{\cal D}) \geq \frac{ H_{\Sigma} }{ 2 }$, where the last inequality follows from  Lemma~\ref{lem:avg_space_LB}, stating that $I_{ \max }( \tau^*) \geq \frac{ H_{\Sigma} }{ 2 }$. Finally, inequality~\eqref{eqn:Nproof_clm_bound_light_3} holds since $\delta = \frac{ \eps^3 }{ 36 \ln (2\Lambda) }$.

\paragraph{Case 2: $\bs{\expar{ \sum_{i \in {\cal L}} Y_i^t } > \frac{2 \eps}{ 3 } \cdot \topt}$.} To arrive at the desired bound, note that since $Y_i^t = \sum_{\tau \in {\cal D}_i} I_{i \tau}^t X_{i\tau}$ and since $\expar{ X_{i\tau} } = \prpar{ \tau^R_i = \tau } = x_{i\tau}^*$, we have
\begin{eqnarray}
&& \pr{ \sum_{i \in {\cal L}} Y_i^t \geq (1 + \eps) \cdot \sum_{i \in {\cal L}} \sum_{\tau \in {\cal D}_i} I_{i \tau}^t x_{i\tau}^* } \nonumber \\
&& \qquad =~~ \pr{ \sum_{i \in {\cal L}} \frac{ Y_i^t }{ \delta H_{\Sigma} } \geq (1 + \eps) \cdot \ex{ \sum_{i \in {\cal L}} \frac{ Y_i^t }{ \delta H_{\Sigma} }} } \nonumber \\
&& \qquad \leq~~ \exp \left( - \frac{ \eps^2 }{ 3\delta } \cdot   \frac{ \expar{ \sum_{i \in {\cal L}} Y_i^t } }{ H_{\Sigma} } \right) \label{eqn:Nproof_clm_bound_light_4} \\
&& \qquad \leq~~ \exp \left( - \frac{ 2\eps^3 }{ 9 \delta } \cdot \frac{ \topt }{ H_{\Sigma} } \right) \label{eqn:Nproof_clm_bound_light_5} \\
&& \qquad \leq~~ \exp \left( - \frac{ \eps^3 }{ 9 \delta }  \right) \label{eqn:Nproof_clm_bound_light_6} \\
&& \qquad \leq~~ \frac{ 1 }{ 2 \Lambda } \ . \label{eqn:Nproof_clm_bound_light_7}
\end{eqnarray}
As in case~1, inequality~\eqref{eqn:Nproof_clm_bound_light_4} follows by specializing~\eqref{eqn:Chernoff-Hoeffding} with respect to $\{ \frac{ Y_i^t }{ \delta H_{\Sigma} } \}_{i \in {\cal L}}$. Inequality~\eqref{eqn:Nproof_clm_bound_light_5} holds due to our case hypothesis. Inequality~\eqref{eqn:Nproof_clm_bound_light_6} is obtained by noting once again that $\topt \geq \frac{ H_{\Sigma} }{ 2 }$. Finally, we arrive at inequality~\eqref{eqn:Nproof_clm_bound_light_7} by plugging in $\delta = \frac{ \eps^3 }{ 36 \ln (2\Lambda) }$.  

%% file: TEX-Approx-Distinct-Intervals.tex
\section{Approximation Scheme in Terms of Distinct Time Intervals} \label{sec:approx_scheme_const_intervals}

In this section, we devise a deterministic approach for discrete inventory staggering, culminating to a polynomial-time approximation scheme in the scenario of constantly-many distinct time intervals. Technically speaking, as stated in  Theorem~\ref{thm:main_result_K}, we will show that for any $\eps \in (0, 1)$, the discrete setting can be approximated within factor $1 + \eps$ of optimal in $O( 2^{ \tilde{O}( K / \eps^2 ) } \cdot | {\cal I} |^{O(1)} )$ time, where $K$ stands for the number of distinct time intervals across all items.

\subsection{\texorpdfstring{$\bs{O ( n^{ O(n) } \cdot | {\cal I} |^{ O(1) } )}$}{}-time peak evaluation oracle} \label{subsec:compute_peak}

As previously mentioned, given an arbitrary shift vector $\tau$, the question of evaluating the peak inventory level $I_{ \max }( {\tau})$ of the replenishment policy ${\cal P}_{\tau}$ in polynomial time is still open. That said, since we will be dealing with  $O( \frac{ K }{ \eps } )$ super-items in Sections~\ref{subsec:mimicking_useful} and~\ref{subsec:mimicking_construct}, designing an exponential-in-$n$ peak evaluation oracle will suffice for our particular purposes.  

\begin{theorem} \label{thm:compute_peak}
For any integer-valued shift vector $\tau$, we can compute the peak inventory level of ${\cal P}_{\tau}$ in $O ( n^{ O(n) } \cdot | {\cal I} |^{ O(1) } )$ time.
\end{theorem}

\paragraph{ILP-formulation.} To derive this result, given a shift vector $\tau \in \bbZ^n$, we assume without loss of generality that $\tau_i \in [0,T_i-1]$ for every item $i \in [n]$. Let us consider the following integer linear program, where our decision variables are $p \in \bbZ$ and $x \in \bbZ^n$: 
\begin{equation} \label{eqn:IP_Imax}
\tag{IP}
\begin{array}{lll}
\max & {\displaystyle \sum_{i \in [n]} H_i \cdot \left( 1 - \frac{ p-(\tau_i + x_i T_i) }{ T_i } \right) \qquad} &  \\
\text{s.t.} & (1)~~ \tau_i + x_i T_i \leq p \qquad & \forall \, i \in [n] \\
& (2)~~ 0 \leq p \leq \Lambda \\
& (3)~~ p \in \bbZ, \, x \in \bbZ^n 
\end{array}
\end{equation}
To develop some basic intuition around this formulation, one can think of $p$ as the time point at which the peak inventory level $I_{ \max }( {\tau})$ is attained. From this perspective, constraints~(2) and~(3) jointly ensure that $p$ is an integer-valued point within $[0, \Lambda]$. Regarding the $x$-variables, for every item $i \in [n]$, the value of $\tau_i + x_i T_i$ represents the last $i$-ordering point in $(-\infty,p]$ with respect to ${\cal P}^i_{  \tau_i }$. As such, the inventory level of this item at time $p$ is given by $H_i \cdot ( 1 - \frac{ p-(\tau_i + x_i T_i) }{ T_i } )$, which is precisely the term we are seeing in the objective function. Along these lines, constraint~(1) guarantees that $\tau_i + x_i T_i$ falls within $(-\infty,p]$, and moreover, since $x_i$ is coupled with a coefficient of $H_i > 0$ in the objective function, an optimal solution will indeed set $\tau_i + x_i T_i$ as the last $i$-ordering point in $(-\infty,p]$.

\paragraph{Basic properties of (\ref{eqn:IP_Imax}).} The next claim explicitly constructs a feasible solution to~\eqref{eqn:IP_Imax} and prescribes lower and upper bounds on the optimum value of this program. 

\begin{lemma} \label{lem:LB_OPT_IP}
\eqref{eqn:IP_Imax} is feasible, and moreover, $\opt\eqref{eqn:IP_Imax} \in [I_{ \max }( {\tau}), H_{ \Sigma }]$.
\end{lemma}
\begin{proof}
To argue about the feasibility of $\eqref{eqn:IP_Imax}$, let us construct a candidate solution $(\hat{p}, \hat{x})$ as follows. First, $\hat{p}$ is chosen as one of the points in $[0, \Lambda] \cap \bbZ$ for which $I_{ \Sigma }( { \tau }, \hat{p}) = I_{ \max }({\tau})$. Next, for every item $i \in [n]$, let $\hat{x}_i = \max \{ x \in \bbZ : \tau_i + x T_i \leq \hat{p} \}$; this choice ensures  that $\tau_i + \hat{x}_i T_i$ is precisely the time point where the last $i$-order in $(-\infty, \hat{p}]$ occurs, with respect to the policy ${\cal P}^i_{ \tau_i }$. Clearly, $(\hat{p}, \hat{x})$ is a feasible solution to~$\eqref{eqn:IP_Imax}$, with an objective value of  
\[ \sum_{i \in [n]} H_i \cdot \left( 1 - \frac{ \hat{p}-(\tau_i + \hat{x}_i T_i) }{ T_i } \right) ~~=~~ \sum_{i \in [n]} I_i( \tau_i, \hat{p}) ~~=~~ I_{ \Sigma }( { \tau }, \hat{p}) ~~=~~ I_{ \max }( {\tau}) \ , \]
implying that we indeed have  $\opt\eqref{eqn:IP_Imax} \geq I_{ \max }( {\tau})$.

In the opposite direction, to prove that $\opt\eqref{eqn:IP_Imax} \leq H_{\Sigma}$, suppose that $(p,x)$ is a feasible solution  to~\eqref{eqn:IP_Imax}. Then, its  objective value can be upper-bounded by noting that
\[ \sum_{i \in [n]} H_i \cdot \left( 1 - \frac{ p-(\tau_i + x_i T_i) }{ T_i } \right) ~~\leq~~ \sum_{i \in [n]} H_i ~~=~~ H_{ \Sigma } \ , \]
where the inequality above holds since $p \geq \tau_i + x_i T_i$, by constraint~(1).
\end{proof}

\paragraph{Inventory peaks via~(\ref{eqn:IP_Imax}).} By Lemma~\ref{lem:LB_OPT_IP}, we know that~\eqref{eqn:IP_Imax} is feasible and bounded, meaning that this formulation indeed has one or more optimal solutions. In what follows, we prove that with respect to any optimal solution $(p^*, x^*)$, the time point $p^*$ must be one where the inventory level of ${\cal P}_{ \tau }$ is maximized.

\begin{lemma} \label{lem:inv_maximizers_IP}
Let $(p^*, x^*)$ be an optimal solution to~\eqref{eqn:IP_Imax}. Then, $I_{ \Sigma }( { \tau }, p^*) = I_{ \max }({\tau})$.
\end{lemma}
\begin{proof}
Suppose on the contrary that $I_{ \Sigma }( { \tau }, p^*) < I_{ \max }({\tau})$, and let us examine the implications of $(p^*, x^*)$ being an optimal  solution to~\eqref{eqn:IP_Imax}. In particular, for every item $i \in [n]$, we must have $x^*_i = \max \{ x \in \bbZ : \tau_i + x T_i \leq p^* \}$. Otherwise, $x^*_i$ can be incremented by $1$, to obtain a feasible solution whose objective value is strictly greater than that of $(p^*, x^*)$. As a result, $\tau_i + x^*_i T_i$ is precisely the time point where the last $i$-order in $(-\infty, p^*]$ occurs, with respect to the policy ${\cal P}^i_{ \tau_i }$, and in turn, $H_i \cdot ( 1 - \frac{ p^*-(\tau_i + x_i^* T_i) }{ T_i } ) = I_i(  \tau_i, p^*)$. We are now ready to reveal the resulting contradiction, arguing that $(p^*, x^*)$ cannot be an optimal solution, by proving that its objective value is  strictly smaller than $\opt\eqref{eqn:IP_Imax}$. To this end, note that   
\begin{eqnarray}
\sum_{i \in [n]} H_i \cdot \left( 1 - \frac{ p^*-(\tau_i + x_i^* T_i) }{ T_i } \right) & = & \sum_{i \in [n]} I_i( \tau_i, p^*) \nonumber \\
& = & I_{\Sigma}( { \tau }, p^*) \nonumber \\
& < & I_{ \max }( {\tau}) \label{eqn:inv_maximizers_IP_1} \\
& \leq & \opt\eqref{eqn:IP_Imax} \ . \label{eqn:inv_maximizers_IP_2}
\end{eqnarray}
Here, inequality~\eqref{eqn:inv_maximizers_IP_1} corresponds to the scenario we are  considering, where $I_{ \Sigma }( \tau, p^*) < I_{ \max }({\tau})$. Inequality~\eqref{eqn:inv_maximizers_IP_2} is precisely our lower bound on $\opt\eqref{eqn:IP_Imax}$, as stated in Lemma~\ref{lem:LB_OPT_IP}.
\end{proof}

\paragraph{Solving~(\ref{eqn:IP_Imax}).} In summary, to compute the peak inventory level of ${\cal P}_{\tau}$, it remains to identify $p^*$, for some optimal solution $(p^*, x^*)$. The next claim, whose proof is provided in Section~\ref{app:proof_lem_solve_IP}, argues that this goal can be accomplished in $O ( n^{ O(n) } \cdot | {\cal I} |^{ O(1) } )$ time. 

\begin{lemma} \label{lem:solve_IP}
For some optimal solution $(p^*, x^*)$ to $\eqref{eqn:IP_Imax}$, we can determine the value of $p^*$  in $O ( n^{ O(n) } \cdot | {\cal I} |^{ O(1) } )$ time.
\end{lemma}

\subsection{Mimicking partitions: Definitions and usefulness} \label{subsec:mimicking_useful}

As previously alluded to, our algorithmic approach will rely on aggregating the overall set of items into a small-sized collection of $O( \frac{ K }{ \eps } )$ super-items, which will be coupled with uniform shifts. For this purpose, let $T_{(1)}, \ldots, T_{(K)}$ be the set of distinct time intervals across $T_1, \ldots, T_n$. For every $k \in [K]$, we use $S_k = \{ i \in [n]: T_i = T_{(k)} \}$ to denote the collection of items associated with a time interval of $T_{(k)}$. Circling back to Section~\ref{subsec:discretize}, we remind the reader that there exists a ${\cal D}$-restricted shift vector $\tau^{\cal D}$ for which $I_{ \max }( {\tau^{\cal D}}) \leq (1 + 4\eps) \cdot \opt^{ \mydisc }( {\cal I} )$. Moreover, for every $k \in [K]$, all items in $S_k$ share the same discretization set, which will be designated by ${\cal D}_{(k)}$, noting that $| {\cal D}_{(k)} | \leq \lfloor \frac{ 1 }{ \eps } \rfloor + 1$.

\paragraph{Mimicking partitions.} For every $k \in [K]$ and $\tau \in {\cal D}_{(k)}$, let $S^{\cal D}_{k, \tau} = \{ i \in S_k : \tau^{\cal D}_i = \tau \}$ be the set of items in $S_k$ whose shift with respect to $\tau^{\cal D}$ is precisely $\tau$. Clearly, the sets $\{ S^{\cal D}_{k, \tau} \}_{k \in [K], \tau \in {\cal D}_{(k)}}$ form a partition of the underlying items, which is  unknown from an algorithmic perspective. To go around this obstacle, we say that a partition $\{ S^{\apx}_{k, \tau} \}_{k \in [K], \tau \in {\cal D}_{(k)}}$ is mimicking $\{ S^{\cal D}_{k, \tau} \}_{k \in [K], \tau \in {\cal D}_{(k)}}$ when it satisfies the next two properties:
\begin{enumerate}
    \item \label{prop_mimick_prop_1} {\em Marginal partition}: $\{ S^{\apx}_{k, \tau} \}_{\tau \in {\cal D}_{(k)}}$ is a partition of $S_k$, for every $k \in [K]$.

    \item \label{prop_mimick_prop_2} {\em Cumulative ordering quantity}: For every $k \in [K]$ and $\tau \in {\cal D}_{(k)}$, 
    \[ H( S^{\apx}_{k, \tau} ) ~~\leq~~ (1 + \eps) \cdot H( S^{\cal D}_{k, \tau} ) + \frac{ 2\eps^2 }{ K } \cdot H_{\Sigma} \ . \]
\end{enumerate}

\paragraph{Usefulness.} To better understand the motivation behind introducing this notion, suppose that $\{ S^{\apx}_{k, \tau} \}_{k \in [K], \tau \in {\cal D}_{(k)}}$ is mimicking $\{ S^{\cal D}_{k, \tau} \}_{k \in [K], \tau \in {\cal D}_{(k)}}$. In this case, let use define a shift vector $\tau^{ \apx }$ such that, for every index $k \in [K]$ and time interval $\tau \in {\cal D}_{(k)}$, all items $i \in S^{\apx}_{k, \tau}$ are given a uniform shift of $\tau^{ \apx }_i = \tau$. The next claim shows that $\tau^{ \apx }$ is near-optimal, meaning that we have just reduced our original inventory staggering problem to the computational question of identifying a mimicking partition.

\begin{lemma} \label{lem:goodness_mimicking}
$I_{ \max }({ \tau^{ \apx } }) \leq (1 + 17\eps) \cdot \opt^{ \mydisc}( {\cal I} )$. 
\end{lemma}
\begin{proof}
To establish the desired claim, consider some index $k \in [K]$, time interval $\tau \in {\cal D}_{(k)}$, and integer-valued point $t \in [0, \Lambda]$. Let $t^-$ be the last $S^{\apx}_{k, \tau}$-order in $(-\infty,t]$ with respect to $\tau^{ \apx }$, namely,  $t^- = \max \{ \tau + m T_{(k)} : \tau + m T_{(k)} \leq t, m \in \bbZ \}$. According to this definition, $I_i( { \tau^{\apx}_i }, t ) = H_i \cdot (1 - \frac{ t - t^- }{ T_{(k)} } )$ for every item $i \in S^{\apx}_{k, \tau}$. Therefore,
\begin{eqnarray}
\sum_{i \in S^{\apx}_{k, \tau}} I_i( { \tau^{ \apx }_i }, t ) & = & \left( 1 - \frac{ t - t^- }{ T_{(k)} } \right) \cdot H ( S^{\apx}_{k, \tau} ) \nonumber \\
& \leq & \left( 1 - \frac{ t - t^- }{ T_{(k)} } \right) \cdot \left( (1 + \eps) \cdot H( S^{\cal D}_{k, \tau} ) + \frac{ 2\eps^2 }{ K } \cdot H_{\Sigma} \right) \label{eqn:goodness_mimicking_1} \\
& \leq & (1 + \eps) \cdot \sum_{i \in S^{\cal D}_{k, \tau}} I_i( { \tau^{\cal D}_i }, t ) + \frac{ 2\eps^2 }{ K } \cdot H_{\Sigma} \ , \label{eqn:goodness_mimicking_2}
\end{eqnarray}
where inequality~\eqref{eqn:goodness_mimicking_1} follows from property~\ref{prop_mimick_prop_2}, and  inequality~\eqref{eqn:goodness_mimicking_2} holds since $I_i( { \tau^{\cal D}_i }, t ) = H_i \cdot (1 - \frac{ t - t^- }{ T_{(k)} } )$ for every item $i \in S^{\cal D}_{k, \tau}$. By summing this relation across all $k \in [K]$ and $\tau \in {\cal D}_{(k)}$, we have
\begin{eqnarray}
I_{ \Sigma }( { \tau^{ \apx } }, t) & = & \sum_{k \in [K]} \sum_{ \tau \in {\cal D}_{(k)} } \sum_{i \in S^{\apx}_{k, \tau}} I_i( { \tau^{ \apx }_i }, t ) \label{eqn:goodness_mimicking_4} \\
& \leq & \sum_{k \in [K]} \sum_{ \tau \in {\cal D}_{(k)} } \left( (1 + \eps) \cdot\sum_{i \in S^{\cal D}_{k, \tau}} I_i( { \tau^{\cal D}_i }, t ) + \frac{ 2\eps^2 }{ K } \cdot H_{\Sigma} \right) \label{eqn:goodness_mimicking_5} \\
& \leq & (1 + \eps) \cdot I_{ \max }( \tau^{\cal D}) + 4\eps  H_{\Sigma} \label{eqn:goodness_mimicking_6} \\
& \leq & (1 + 17\eps) \cdot \opt^{ \mydisc}( {\cal I} ) \ . \label{eqn:goodness_mimicking_7}
\end{eqnarray}
Here, equality~\eqref{eqn:goodness_mimicking_4} holds since $\{ S^{\apx}_{k, \tau} \}_{k \in [K], \tau \in {\cal D}_{(k)}}$ is a partition of the entire item set, by property~\ref{prop_mimick_prop_1}. Inequality~\eqref{eqn:goodness_mimicking_5} follows by plugging in~\eqref{eqn:goodness_mimicking_2}. Inequality~\eqref{eqn:goodness_mimicking_6} is obtained by recalling that 
$| {\cal D}_{(k)} | \leq \lfloor \frac{ 1 }{ \eps } \rfloor + 1 \leq \frac{ 2 }{ \eps }$. Finally, to arrive at inequality~\eqref{eqn:goodness_mimicking_7}, note that $I_{ \max }({\tau^{\cal D}}) \leq (1 + 4\eps) \cdot \opt^{ \mydisc}( {\cal I} )$ and that $H_{ \Sigma } \leq 2 \cdot \opt^{ \mydisc}( {\cal I} )$, by Lemma~\ref{lem:avg_space_LB}.
\end{proof}

\subsection{Mimicking partitions: Construction and testing} \label{subsec:mimicking_construct}

In what follows, we construct a family of $2^{ O( K / \eps^2 ) }$ partitions, out of which at least one is guaranteed to be mimicking $\{ S^{\cal D}_{k, \tau} \}_{k \in [K], \tau \in {\cal D}_{(k)}}$; this construction can be implemented in $O( 2^{ O( K / \eps^2 ) } \cdot n^{O(1)} )$ time. Subsequently, we will explain how to utilize this family in order to compute a near-optimal shift vector.

\paragraph{Step 1: Guessing.} As our first algorithmic step, for every $k \in [K]$ and $\tau \in {\cal D}_{(k)}$, we guess an over-estimate $H^{\apx}_{k, \tau}$ for $H( S^{\cal D}_{k, \tau} )$, such that 
\begin{equation} \label{eqn:ineq_guess_Hapx}
H( S^{\cal D}_{k, \tau} ) ~~<~~ H^{\apx}_{k, \tau} ~~\leq~~ H( S^{\cal D}_{k, \tau} ) + \frac{ \eps^2 }{ K } \cdot H_{\Sigma} \ .
\end{equation}
To this end, it suffices to enumerate over integer multiples of $\frac{ \eps^2 }{ K } \cdot H_{\Sigma}$ within the interval $[0, H_{\Sigma}]$. Based on elementary balls-into-bins arguments, across all combinations of $k \in [K]$ and $\tau \in {\cal D}_{(k)}$, the required number of guesses is $2^{ O( K / \eps^2 ) }$. 

\paragraph{Step 2: Load balancing formulation.} For every $k \in [K]$, we proceed by separately partitioning the set of items $S_k$ into $\{ S^{\apx}_{k, \tau} \}_{\tau \in {\cal D}_{(k)}}$. For this purpose, we create the following instance of makespan minimization on related (uniform)  machines:
\begin{itemize}
    \item {\em Jobs}: Each item $i \in S_k$ corresponds to a job, requiring $H_i$ amount of work.

    \item {\em Machines}: For every $\tau \in {\cal D}_{(k)}$, we represent the subset $S^{\apx}_{k, \tau}$ via a corresponding machine ${\cal M}_{k, \tau}$, whose speed is $H^{\apx}_{k, \tau}$. As such, when job $i$ is assigned to this machine, it incurs a processing time of $\frac{ H_i }{ H^{\apx}_{k, \tau} }$.
    
    \item {\em Goal}: Our objective is to compute a job-to-machine assignment whose makespan is minimized.    
\end{itemize}
It is easy to see that the optimal makespan for this particular instance is at most $1$. Indeed, when for every $\tau \in {\cal D}_{(k)}$, the unknown set of jobs $S^{\cal D}_{k, \tau}$ is assigned to machine ${\cal M}_{k, \tau}$, the latter incurs a total processing time of $\frac{ H( S^{\cal D}_{k, \tau} ) }{ H^{\apx}_{k, \tau} } < 1$, due to inequality~\eqref{eqn:ineq_guess_Hapx}.

\paragraph{Step 3: Approximating the load balancing problem.} Luckily, makespan minimization on related machines admits an efficient polynomial-time approximation scheme (EPTAS). Specifically, \citet{JansenKV20} showed how to compute an assignment whose makespan is within factor $1 + \eps$ of optimal; their approach can be implemented in $O( 2^{ \tilde{O}( 1/\eps ) } + n^{O(1)} )$ time. This result improves on a slightly less efficient approach by \citet{Jansen10}. Migrated to our context, it follows that within this running time, we can partition $S_k$ into $\{ S^{\apx}_{k, \tau} \}_{\tau \in {\cal D}_{(k)}}$, such that for every $\tau \in {\cal D}_{(k)}$,
\[ H( S^{\apx}_{k, \tau} ) ~~\leq~~ (1 + \eps) \cdot H^{\apx}_{k, \tau} ~~\leq~~ (1 + \eps) \cdot H( S^{\cal D}_{k, \tau} ) + \frac{ 2\eps^2 }{ K } \cdot H_{\Sigma} \ , \]
where the last transition follows from inequality~\eqref{eqn:ineq_guess_Hapx}. In other words, $\{ S^{\apx}_{k, \tau} \}_{k \in [K], \tau \in {\cal D}_{(k)}}$ is a mimicking partition for $\{ S^{\cal D}_{k, \tau} \}_{k \in [K], \tau \in {\cal D}_{(k)}}$.

\paragraph{Testing.} Up until now, we have constructed a family of $2^{ O( K / \eps^2 ) }$ partitions, containing at least one which is guaranteed to be mimicking $\{ S^{\cal D}_{k, \tau} \}_{k \in [K], \tau \in {\cal D}_{(k)}}$. As shown in Section~\ref{subsec:mimicking_useful}, the shift vector $\tau^{ \apx }$ corresponding to such a partition, $\{ S^{\apx}_{k, \tau} \}_{k \in [K], \tau \in {\cal D}_{(k)}}$, forms a $(1 + 17\eps)$-approximation. Therefore, the remaining question is: Given $\tau^{ \apx }$, how do we compute its peak inventory level, $I_{ \max }({ \tau^{ \apx } })$?

To resolve this question, we remind the reader that $\tau^{ \apx }$ is defined such that, for every index $k \in [K]$ and time interval $\tau \in {\cal D}_{(k)}$, all items in $S^{\apx}_{k, \tau}$ are given a uniform shift of $\tau$. Thus, to evaluate $I_{ \max }({ \tau^{ \apx } })$, we can view $S^{\apx}_{k, \tau}$ as a super-item, $(k,\tau)$, associated with a time interval of $T_{(k)}$ and with an ordering quantity of $H( S^{\apx}_{k, \tau} )$. Consequently, it remains to compute the peak inventory level of a given shift vector, when the underlying number of items is $O( \frac{ K }{ \eps } )$. By Theorem~\ref{thm:compute_peak}, $I_{ \max }({ \tau^{ \apx } })$ can be evaluated in $O ( (\frac{ K }{ \eps })^{ O(K/\eps) } \cdot | {\cal I} |^{ O(1) } ) = O ( 2^{ \tilde{O}(K/\eps) } \cdot | {\cal I} |^{ O(1) } )$ time.

\subsection{Proof of Lemma~\ref{lem:solve_IP}} \label{app:proof_lem_solve_IP}

\paragraph{Integer linear feasibility.} For the purpose of computing an optimal solution to~\eqref{eqn:IP_Imax}, we will rely on the work of \citet{Lenstra83}, \citet{Kannan87}, and \citet{FrankT87} for solving integer linear feasibility problems in fixed dimension. Specifically, given an integer-valued matrix $A \in \bbZ^{m \times n}$ along with an integer-valued vector $b \in \bbZ^m$, suppose we wish to decide whether there exists a vector $x \in \bbZ^n$ satisfying $Ax \leq b$. Then, the above-mentioned papers aggregately proved that this problem is decidable in $O( n^{ O(n) } \cdot L_{A,b} )$ time, where $L_{A,b}$ represents the combined number of bits required to specify $A$ and $b$.

\paragraph{The feasibility version of~(\ref{eqn:IP_Imax}).} Prior to proceeding any further, we mention in passing that the upcoming procedure for optimizing~\eqref{eqn:IP_Imax} is rather standard, and we refer interested readers to two excellent sources --- \citet[Sec.~2.8]{Lokshtanov09} and \citet[Sec.~6.2]{CyganFKLMPPS15} --- for additional explanations about obtaining an optimal solution to a given integer linear program via its feasibility counterpart. In order to  convert~\eqref{eqn:IP_Imax} into a feasibility problem of the form described above, rather than having $\sum_{i \in [n]} H_i \cdot ( 1 - \frac{ p-(\tau_i + x_i T_i) }{ T_i } )$ as our objective function, when this term is scaled by a factor of $\Lambda$ and lower-bounded by $\psi$, we arrive at the next integer linear feasibility problem:
\begin{equation} \label{eqn:IP_feasible}
\tag{IP$_{\text{feasible}}$}
\begin{array}{ll}
(1)~~ {\displaystyle \Lambda \cdot \sum_{i \in [n]} H_i \cdot \left( 1 - \frac{ p-(\tau_i + x_i T_i) }{ T_i } \right) \geq \psi \qquad} \\
(2)~~ \tau_i + x_i T_i \leq p \qquad & \forall \, i \in [n] \\
(3)~~ 0 \leq p \leq \Lambda \\
(4)~~ p \in \bbZ, \, x \in \bbZ^n 
\end{array}
\end{equation}

\paragraph{Optimizing (\ref{eqn:IP_Imax}) via~(\ref{eqn:IP_feasible}).} Let us observe that, as long as $\psi$ is an integer, the constraint matrix $A$ of this formulation and its right-side vector $b$ are integer-valued. Therefore, subsequently to the opening paragraph of this section, we know that~\eqref{eqn:IP_feasible} can be decided in $O( n^{ O(n) } \cdot L_{A,b,\psi} )$ time. Here, the absolute value of every entry in $A$ and $b$ is at most $\Lambda n H_{\max}\psi$, with room to spare, implying that $L_{A,b,\psi} = O( n^2 \log (\Lambda n H_{\max}\psi) )$. 

Now, by Lemma~\ref{lem:LB_OPT_IP}, we know that $\opt\eqref{eqn:IP_Imax} \in [I_{ \max }({\tau}), H_{ \Sigma }] \subseteq [\frac{ H_{\Sigma} }{ 2 }, H_{ \Sigma }]$. During the construction of~\eqref{eqn:IP_feasible}, we have scaled our objective function by $\Lambda$, meaning that the exact optimum of~\eqref{eqn:IP_Imax} can be identified via binary search over the interval $[\frac{ \Lambda H_{\Sigma} }{ 2 }, \Lambda H_{ \Sigma }]$. This search will consist of $O( \log ( \Lambda H_{\Sigma} ) )$ iterations, each employing a single call to~\eqref{eqn:IP_feasible} with a threshold of $\psi = \Theta( \Lambda H_{\Sigma} )$. As a result, $\opt\eqref{eqn:IP_Imax}$ can be computed in time 
\[ O \left( n^{ O(n) } \cdot  \log^2 (\Lambda H_{\max} ) \right) ~~=~~ O \left( n^{ O(n) } \cdot  \log^2 (T_{\max} H_{\max} ) \right) ~~=~~ O \left( n^{ O(n) } \cdot | {\cal I} |^{ O(1) } \right) \ . \]

Next, our objective is to determine the value of $p^*$, for some  optimal solution $(p^*, x^*)$ to $\eqref{eqn:IP_Imax}$. For this purpose, we augment~\eqref{eqn:IP_feasible} with two constraints:
\begin{itemize}
    \item First, in place of $\Lambda \cdot \sum_{i \in [n]} H_i \cdot ( 1 - \frac{ p-(\tau_i + x_i T_i) }{ T_i } ) \geq \psi$, since $\opt\eqref{eqn:IP_Imax}$ is already known, we focus on optimal solutions by writing $\Lambda \cdot \sum_{i \in [n]} H_i \cdot ( 1 - \frac{ p-(\tau_i + x_i T_i) }{ T_i } )= \Lambda \cdot \opt\eqref{eqn:IP_Imax}$.

    \item Our second constraint is of the form $p \geq \psi$, intended to conduct a binary search for the value of $p^*$.
\end{itemize}
Similarly to the line of reasoning described above, $p^*$ can be computed in $O ( n^{ O(n) } \cdot | {\cal I} |^{ O(1) } )$ time as well. 

%% file: TEX-Approx-Nested.tex
\section{Approximation Scheme for  Nested Instances} \label{sec:approx_scheme_nested}

In this section, we develop a polynomial-time approximation scheme for nested instances of the inventory staggering problem, thereby establishing Theorem~\ref{thm:main_result_nested}. Specifically, for any $\eps > 0$, our decomposition-based approach will approximate such instances within factor $1 + \eps$ of optimal in $O ( 2^{ \tilde{O}(1/\eps^3) } \cdot | {\cal I} |^{ O(1) } )$ time. For ease of exposition, these ideas will be described in terms of the discrete problem formulation; we will therefore open by explaining how  continuous instances can  be reduced to discrete ones, with negligible loss in optimality. 

\subsection{Continuous-to-discrete reduction} \label{subsec:reduction_cont_disc}

Interestingly, the discretization ideas presented in Section~\ref{subsec:discretize} allow us to reduce any instance ${\cal I}_{\bbR^n}$ of the continuous model to a discrete instance, ${\cal I}_{\bbZ^n}$, losing a factor of $1 + O(\eps)$ in its objective value. For this purpose, all time intervals $T_1, \ldots, T_n$ are uniformly scaled up by a factor of $\lceil \frac{ 1 }{ \eps } \rceil$ and nothing more. This modification clearly preserves nestedness.

To analyze the relationship between ${\cal I}_{\bbR^n}$ and ${\cal I}_{\bbZ^n}$, let us first observe that due to uniformly up-scaling all time intervals, when scaled down by a factor of $\lceil \frac{ 1 }{ \eps } \rceil$, the peak inventory level of any integer-valued shift vector $\tau$ for ${\cal I}_{\bbZ^n}$ translates to precisely the same peak for ${\cal I}_{\bbR^n}$. In other words, $I_{\max}^{ {\cal I}_{\bbR^n} }( \frac{ \tau }{ \lceil 1/\eps \rceil } ) = I_{\max}^{ {\cal I}_{\bbZ^n} }( \tau )$, implying that $\opt^{ \mycont }( {\cal I}_{\bbR^n} ) \leq \opt^{ \mydisc }( {\cal I}_{\bbZ^n} )$. To relate these measures in the opposite direction, let $\tilde{T}_1, \ldots, \tilde{T}_n$ be the time intervals of ${\cal I}_{\bbZ^n}$, and suppose we associate each item $i \in [n]$ with a discrete set of  integer-valued shifts, ${\cal D}_i = \{ \lfloor \eps \tilde{T}_i \rfloor, \lfloor 2\eps \tilde{T}_i \rfloor, \ldots, \lfloor \lfloor \frac{ 1 }{ \eps } \rfloor \cdot \eps \tilde{T}_i \rfloor , \tilde{T}_i \}$. Then, minor modifications to the proof of Lemma~\ref{lem:D_restricted_good} show that there exists a ${\cal D}$-restricted shift vector for ${\cal I}_{\bbZ^n}$ whose peak inventory level is $(1 + O(\eps)) \cdot \opt^{ \mycont }( {\cal I}_{\bbR^n} )$.

\subsection{Well-separated partitions}

Our approach for devising a polynomial-time approximation scheme is driven by the notion of well-separated partitions. To gradually introduce these objects, given a nested instance ${\cal I}$ of the discrete inventory staggering problem, suppose that  ${\cal S} = (S_1, \ldots, S_M, S_{\infty})$ is an ordered partition of its item set. Then, ${\cal S}$ is called well-separated when it satisfies the next three properties:
\begin{enumerate}
    \item \label{prop:well_separated_2} {\em Width}: $\frac{ \max_{i \in S_m} T_i }{ \min_{i \in S_m} T_i } \leq (\frac{ 1 }{ \eps })^{ 1/\eps }$, for every $m \in [M]$.

    \item \label{prop:well_separated_3} {\em Separation}: $T_{ i_1 } \leq \eps T_{i_2}$, for every $(i_1, i_2) \in S_{m_1} \times S_{m_2}$ with $1 \leq m_1 < m_2 \leq M$.  

    \item \label{prop:well_separated_1} {\em $S_{\infty}$ is negligible}: $H( S_{\infty} )  \leq \eps H_{\Sigma}$.
\end{enumerate}
In what follows, we prove the existence of well-separated partitions and propose an efficient construction through a simple application of the probabilistic method \citep{AlonS2016}. 

\begin{lemma}
A well-separated partition exists, and can be deterministically computed in polynomial time.    
\end{lemma}
\begin{proof}
We begin by defining the sequence of time interval subsets 
\[ {\cal T}_1 ~~=~~ \left\{ i \in [n] : T_i \in \left[ 1, \frac{ 1 }{ \eps } \right] \right\}, \qquad {\cal T}_2 ~~=~~ \left\{ i \in [n] : T_i \in \left( \frac{ 1 }{ \eps }, \frac{ 1 }{ \eps^2 } \right] \right\}, \qquad \ldots \]
so on and so forth, where in general ${\cal T}_q = \left\{ i \in [n] : T_i \in \left( \frac{ 1 }{ \eps^{q-1} }, \frac{ 1 }{ \eps^q } \right] \right\}$ for every $q \geq 2$. This sequence proceeds up to ${\cal T}_Q$, where $Q$ is the smallest integer for which $\frac{ 1 }{ \eps^Q } \geq T_{\max}$, meaning that $Q = \lceil \log_{ 1/\eps } T_{\max} \rceil = O(  \log T_{\max} )$.

Assuming without loss of generality that $\frac{ 1 }{ \eps }$ is an integer, we randomly pick $\xi \sim U \{ 0, \ldots, \frac{ 1 }{ \eps } - 1 \}$. With respect to this choice, any subset ${\cal T}_q$ with $q \equiv \xi \ (\mymod \ \frac{ 1 }{ \eps })$ is said to be marked, while all other subsets are unmarked. We use $S_1^{ \xi }$ to denote the collection of time intervals belonging to the (unmarked) subsets appearing prior to the first marked subset (i.e., ${\cal T}_1, \ldots, {\cal T}_{\xi-1}$). Then, $S_2^{ \xi }$ will denote the collection of time intervals in the (unmarked) subsets appearing between the first and second marked subsets (i.e., ${\cal T}_{\xi+1}, \ldots, {\cal T}_{\xi + \frac{ 1 }{ \eps } - 1}$), so on and so forth, up until the last such set, $S_M^{ \xi }$. Finally, time intervals within marked subsets are collected into $S_{\infty}^{ \xi }$.

It is easy to verify that, for any choice of $\xi$, the ordered partition ${\cal S}^{\xi} = (S_1^{ \xi }, \ldots, S_M^{ \xi }, S_{\infty}^{ \xi })$ satisfies properties~\ref{prop:well_separated_2}  and~\ref{prop:well_separated_3}. In addition, 
\[ \ex{ H(  S^{\xi}_{\infty} ) } ~~=~~ \ex{ \sum_{i \in [n]} H_i \cdot 1 [ i \in S^{\xi}_{\infty} ] } ~~=~~ \eps H_{\Sigma} \ , \]
since each item $i \in [n]$ ends up in $S^{\xi}_{\infty}$ if and only if the unique subset ${\cal T}_q$ containing this item is marked, which happens with probability $\eps$. As an immediate consequence, it follows that there exists some choice of $\xi \in \{ 0, \ldots, \frac{ 1 }{ \eps } - 1 \}$ for which $H(  S^{\xi}_{\infty} ) \leq \eps H_{\Sigma}$, meaning that property~\ref{prop:well_separated_1} holds as well.
\end{proof}

\subsection{The Near-Additivity Theorem} \label{subsec:near_additive_nested}

Given a well-separated partition ${\cal S} = (S_1, \ldots, S_M, S_{\infty})$, our next step consists of arguing that, by temporarily leaving the subset $S_{\infty}$ aside, we are left with a residual instance whose optimal peak inventory level is nearly additive with respect to $S_1, \ldots, S_M$.  To formalize this claim, for every $m \in [M]$, we make use of ${\cal I}_m$ to denote the instance obtained by restricting  ${\cal I}$ to the set of items $S_m$. Similarly, ${\cal I}_{[m]}$ will designate its restriction to $S_{[m]} = \bigcup_{\mu \in [m]} S_{\mu}$. With this notation, it is easy to verify that $\opt^{ \mydisc }( {\cal I}_{[M]} ) \leq \sum_{m \in [M]} \opt^{ \mydisc }( {\cal I}_m )$. However, we will show that well-separated partitions, in conjunction with nested time intervals, lead to a nearly-tight inequality in the opposite direction.

\begin{theorem} \label{thm:near_additive_THM}
$\opt^{ \mydisc }( {\cal I}_{[M]} ) \geq (1 - 2\eps) \cdot \sum_{m \in [M]} \opt^{ \mydisc }( {\cal I}_m )$.
\end{theorem}

\paragraph{The basic claim.} In what follows, we argue that for every $m \in [M]$,
\begin{equation} \label{eqn:NA_Theorem_induction}
\opt^{ \mydisc }( {\cal I}_{[m]} ) ~~\geq~~ \opt^{ \mydisc }( {\cal I}_{[m-1]} ) + (1 - 2\eps) \cdot \opt^{ \mydisc }( {\cal I}_m ) \ ,
\end{equation}
with the convention that $\opt^{ \mydisc }( {\cal I}_{[0]} ) = 0$. By expanding this inequality one term after the other, we indeed get $\opt^{ \mydisc }( {\cal I}_{[M]} ) \geq (1 - 2\eps) \cdot \sum_{m \in [M]} \opt^{ \mydisc }( {\cal I}_m )$, as desired. Noting that inequality~\eqref{eqn:NA_Theorem_induction} is trivial for $m=1$, we proceed by considering the general case of $m \geq 2$.

\paragraph{The $\bs{\eps}$-shortness relation.} To this end, suppose that ${\cal I}_{\cal A}$ and ${\cal I}_{\cal B}$ are two inventory staggering instances, whose corresponding sets of items ${\cal A}$ and ${\cal B}$ are disjoint. For $\eps \in (0,1)$, we say that ${\cal I}_{\cal A}$ is $\eps$-shorter than ${\cal I}_{\cal B}$ when $\lcm( \{ T_i \}_{i \in {\cal A}} ) \leq \eps \cdot \min_{i \in {\cal B}} T_i$. Put differently, the cycle length of any policy in regard to ${\cal I}_{\cal A}$ is at most an $\eps$-fraction of the minimal time interval in ${\cal I}_{\cal B}$. The next claim shows that, in this scenario, the best-possible peak inventory level for the merged instance ${\cal I}_{{\cal A} \cup {\cal B}}$ cannot be much smaller than the sum of optimal peaks for ${\cal I}_{\cal A}$ and ${\cal I}_{\cal B}$. It is important to point out that this result applies to both versions of the inventory staggering problem, which is why we make use of  $\opt(\cdot)$, rather than $\opt^{ \mydisc }(\cdot)$ or $\opt^{ \mycont }(\cdot)$.

\begin{lemma} \label{lem:eps_shorter_additive}
When ${\cal I}_{\cal A}$ is $\eps$-shorter than ${\cal I}_{\cal B}$, 
\[ \opt( {\cal I}_{{\cal A} \cup {\cal B}} ) ~~\geq~~ \opt( {\cal I}_{\cal A}) + (1 - 2\eps) \cdot \opt( {\cal I}_{\cal B} ) \ . \] 
\end{lemma}
\begin{proof}
Let $\tau^{*{\cal A} \cup {\cal B}}$ be an optimal shift vector with respect to ${\cal I}_{{\cal A} \cup {\cal B}}$. In terms of this vector, the total inventory level of ${\cal B}$-items at any time point $t \in \bbR$ will be denoted by $I_{ \Sigma }^{{\cal B}}( \tau^{*{\cal A} \cup {\cal B}} , t )  = \sum_{i \in {\cal B}} I_i( \tau^{*{\cal A} \cup {\cal B}}_i , t )$, with its peak level being $I_{ \max }^{\cal B}( \tau^{*{\cal A} \cup {\cal B}} ) = \max_{t \in \bbR} I_{ \Sigma }^{{\cal B}}( \tau^{*{\cal A} \cup {\cal B}} , t )$. Now, suppose that the latter $\max$ is attained at $\hat{t}$, and let $\Delta = \eps T^{\cal B}_{\min}$, where $T^{\cal B}_{\min} = \min_{i \in {\cal B}} T_i$. Our first claim is that, for every time point $t \in [\hat{t}, \hat{t} + \Delta]$, we have 
\begin{equation} \label{eqn:proof_near_additive_Im}
I_{ \Sigma }^{{\cal B}}( \tau^{*{\cal A} \cup {\cal B}} , t ) ~~\geq~~ (1 - 2\eps) \cdot I_{ \max }^{\cal B}( \tau^{*{\cal A} \cup {\cal B}} ) \ .
\end{equation}
For this purpose, note that when there are no ${\cal B}$-orders in $[\hat{t},t]$,
\begin{eqnarray}
I_{ \Sigma }^{{\cal B}}( \tau^{*{\cal A} \cup {\cal B}} , t ) & = &I_{ \Sigma }^{{\cal B}}( \tau^{*{\cal A} \cup {\cal B}} , \hat{t} ) - (t - \hat{t}) \cdot \sum_{i \in {\cal B}} \frac{ H_i }{ T_i } \nonumber \\
& \geq & I_{ \max }^{\cal B}( \tau^{*{\cal A} \cup {\cal B}} ) - \eps T^{\cal B}_{\min} \cdot \sum_{i \in {\cal B}} \frac{ H_i }{ T_i } \label{eqn:proof_near_additive_THM_1} \\
& \geq & I_{ \max }^{\cal B}( \tau^{*{\cal A} \cup {\cal B}} ) - \eps \cdot H( {\cal B} ) \nonumber \\
& \geq & (1 - 2\eps) \cdot I_{ \max }^{\cal B}( \tau^{*{\cal A} \cup {\cal B}} ) \ . \label{eqn:proof_near_additive_THM_2}
\end{eqnarray}
Here, inequality~\eqref{eqn:proof_near_additive_THM_1} holds since $I_{ \Sigma }^{{\cal B}}( \tau^{*{\cal A} \cup {\cal B}} , \hat{t} ) = I_{ \max }^{\cal B}( \tau^{*{\cal A} \cup {\cal B}} )$, by definition of $\hat{t}$, and since $t - \hat{t} \leq \Delta = \eps T^{\cal B}_{\min}$. Inequality~\eqref{eqn:proof_near_additive_THM_2} is obtained by specializing Lemma~\ref{lem:avg_space_LB} to the set of items ${\cal B}$, implying that $H( {\cal B} ) \leq 2 \cdot \opt( {\cal I}_ {\cal B} ) \leq 2 \cdot I_{ \max }^{\cal B}( \tau^{*{\cal A} \cup {\cal B}} )$. In the complementary case, where there are one or more ${\cal B}$-orders in $[\hat{t},t]$, the inventory level $I_{ \Sigma }^{{\cal B}}( \tau^{*{\cal A} \cup {\cal B}} , t )$ can only be larger.

Now, let us recall that ${\cal I}_{\cal A}$ is $\eps$-shorter than ${\cal I}_{\cal B}$, meaning that the cycle length of any policy with respect to ${\cal I}_{\cal A}$ is at most $\eps T^{\cal B}_{\min} = \Delta$. In particular, focusing on the restriction of $\tau^{*{\cal A} \cup {\cal B}}$ to these items, within the interval $[\hat{t}, \hat{t} + \Delta]$, there exists at least one point $\bar{t}$ with
\begin{equation} \label{eqn:proof_near_additive_I_1_m-1} I_{ \Sigma }^{\cal A}( \tau^{*{\cal A} \cup {\cal B}}, \bar{t}) ~~=~~ I_{\max}^{\cal A}( \tau^{*{\cal A} \cup {\cal B}}) \ .
\end{equation}
Consequently, to derive the desired lower bound on $\opt( {\cal I}_{{\cal A} \cup {\cal B}} )$, we observe that
\begin{eqnarray} 
\opt( {\cal I}_{{\cal A} \cup {\cal B}} ) & = & I_{\max}^{{\cal A} \cup {\cal B}}(\tau^{*{\cal A} \cup {\cal B}}) \nonumber \\
& \geq & I_{\Sigma}^{{\cal A} \cup {\cal B}}( \tau^{*{\cal A} \cup {\cal B}}, \bar{t}) \nonumber \\
& = & I_{\Sigma}^{{\cal A}}( \tau^{*{\cal A} \cup {\cal B}}, \bar{t}) + I_{\Sigma}^{ {\cal B}}( \tau^{*{\cal A} \cup {\cal B}}, \bar{t}) \nonumber \\
& \geq & I_{\max}^{\cal A}( \tau^{*{\cal A} \cup {\cal B}}) + 
(1 - 2\eps) \cdot I_{ \max }^{\cal B}( \tau^{*{\cal A} \cup {\cal B}} ) \label{eqn:proof_near_additive_THM_3} \\
& \geq & \opt( {\cal I}_{\cal A}) + (1 - 2\eps) \cdot \opt( {\cal I}_{\cal B} ) \ , \nonumber
\end{eqnarray}
where inequality~\eqref{eqn:proof_near_additive_THM_3} is obtained by plugging in~\eqref{eqn:proof_near_additive_Im} and~\eqref{eqn:proof_near_additive_I_1_m-1}.    
\end{proof}

\paragraph{Concluding the basic claim~(\ref{eqn:NA_Theorem_induction}).} As our last step, we explain how to derive inequality~\eqref{eqn:NA_Theorem_induction} for the general case of $m \geq 2$. To this end, let us recall that ${\cal S}$ is a well-separated partition, meaning that for every pair of items $i_1 \in S_{[m-1]}$ and $i_2 \in S_m$, we necessarily have $T_{i_1} \leq \eps T_{i_2}$, by property~\ref{prop:well_separated_3}. Consequently, since we are considering a nested instance, the cycle length of any policy with respect to ${\cal I}_{[m-1]}$ is given by $\max_{ i_1 \in S_{[m-1]} } T_{i_1} \leq \eps \cdot \min_{ i_2 \in S_m } T_{i_2}$, implying that ${\cal I}_{[m-1]}$ is $\eps$-shorter than ${\cal I}_{m}$. As a result, by Lemma~\ref{lem:eps_shorter_additive},
\begin{eqnarray*} 
\opt^{ \mydisc }( {\cal I}_{[m]} ) & \geq & \opt^{ \mydisc }( {\cal I}_{[m-1]} ) + (1 - 2\eps) \cdot \opt^{ \mydisc }( {\cal I}_m ) \ . 
\end{eqnarray*}

\subsection{Algorithmic implications}

From an algorithmic perspective, in light of the preceding discussion, we first  compute a $(1+\eps)$-approximate shift vector $\tau^m$ with respect to ${\cal I}_m$, for every $m \in [M]$. To this end, by property~\ref{prop:well_separated_2}, we know that $\frac{ \max_{i \in S_m} T_i }{ \min_{i \in S_m} T_i } \leq (\frac{ 1 }{ \eps })^{ 1/\eps }$. Additionally, since ${\cal I}_m$ is a nested instance, the multiplicative gap between any pair of distinct time intervals in $\{ T_i \}_{i \in S_m}$ is at least $2$, implying that there are only $O( \log_2 ((\frac{ 1 }{ \eps })^{ 1/\eps }) ) = O( \frac{ 1 }{ \eps } \log \frac{ 1 }{ \eps } )$ such values. Therefore, as stated in Theorem~\ref{thm:main_result_K}, we can compute $\tau^m$ in $O ( 2^{ \tilde{O}(1/\eps^3) } \cdot | {\cal I} |^{ O(1) } )$ time. For the singular instance ${\cal I}_{\infty}$, we pick an arbitrary shift vector $\tau^{ \infty }$. Our final solution $\tau$ is formed by gluing $\tau^1, \ldots, \tau^M, \tau^{ \infty }$ together. The next claim proves that this vector is indeed near-optimal in terms of the original instance ${\cal I}$.

\begin{lemma}
$I_{\max}( \tau ) \leq (1 + 8\eps) \cdot \opt^{ \mydisc }( {\cal I} )$.   
\end{lemma}
\begin{proof}
To upper-bound the peak inventory level of the policy ${\cal P}_{\tau}$, we observe that 
\begin{eqnarray}
I_{\max}( \tau ) & \leq & \sum_{m \in [M]} I_{\max}( \tau^m ) + I_{\max}( \tau^{ \infty } ) \nonumber \\
& \leq & (1 + \eps) \cdot \sum_{m \in [M]} \opt^{ \mydisc }( {\cal I}_m ) + H( S_{\infty} ) \label{eqn:alg_imp_nested_1} \\
& \leq & \frac{ 1 + \eps }{ 1 - 2\eps } \cdot \opt^{ \mydisc }( {\cal I}_{[M]} ) + H( S_{\infty} ) \label{eqn:alg_imp_nested_2} \\
& \leq & \left( \frac{ 1 + \eps }{ 1 - 2\eps } + 2\eps \right) \cdot \opt^{ \mydisc }( {\cal I} ) \label{eqn:alg_imp_nested_3} \\
& \leq & ( 1 + 8\eps ) \cdot \opt^{ \mydisc }( {\cal I} ) \ . \nonumber
\end{eqnarray}
Here, inequality~\eqref{eqn:alg_imp_nested_1} holds since $\tau^m$ is a $(1+\eps)$-approximate vector with respect to ${\cal I}_m$, for every $m \in [M]$. Inequality~\eqref{eqn:alg_imp_nested_2} follows from Theorem~\ref{thm:near_additive_THM}. Finally, inequality~\eqref{eqn:alg_imp_nested_3} is obtained by observing that $H ( S_{\infty} ) \leq \eps H_{\Sigma} \leq 2\eps \cdot \opt^{ \mydisc }( {\cal I} )$, due to  property~\ref{prop:well_separated_1} and Lemma~\ref{lem:avg_space_LB}.   
\end{proof}

%% file: TEX-Approx-Coprime.tex
\section{Approximation Scheme for Continuous Pairwise Coprime Instances} \label{sec:approx_comprime}

Our final contribution comes in the form of a polynomial-time approximation scheme for pairwise coprime instances of the continuous inventory staggering problem. As stated in Theorem~\ref{thm:main_result_coprime}, by repurposing some of the main ideas behind Section~\ref{sec:approx_scheme_nested}, we will approximate such instances within factor $1 + \eps$ of optimal in $O( \mytower_4(O(\frac{ 1 }{ \eps}), O(\frac{ 1 }{ \eps})) \cdot | {\cal I} |^{O(1)} )$ time.   

\subsection{Instilling \texorpdfstring{$\bs{\eps}$}{}-shortness and near-additivity}

\paragraph{Decomposition.} Given an instance ${\cal I}$ of the continuous inventory staggering problem, comprised of pairwise coprime time intervals $T_1 < \cdots < T_n$, our first objective is to decompose ${\cal I}$ into two essentially independent instances, with one being $\eps$-shorter than the other. For this purpose, assuming without loss of generality that $\frac{ 1 }{ \eps }$ takes an integer value, we define the infinite sequence  $\Psi_1, \Psi_2, \ldots$, such that  
\[ \Psi_0 ~~=~~ 1, \qquad \Psi_1 ~~=~~ \frac{ 1 }{ \eps}, \qquad \Psi_2 ~~=~~ 4^{ 1/\eps }, \qquad \Psi_3 ~~=~~ 4^{ \displaystyle 4^{ 1/\eps }}, \qquad \ldots \]
where in general, $\Psi_m = \mytower_4(m, \frac{ 1 }{ \eps})$ for all $m \geq 1$. In correspondence to this sequence, we additionally define the items sets $S_1, S_2, \ldots$, where $S_m = \{ i \in [n]: T_i \in [\Psi_{m-1}, \Psi_m) \}$. Clearly, there exists some $2 \leq \hat{m} \leq \frac{ 1 }{ \eps}+1$ for which $H( S_{\hat{m}} ) \leq \eps H_{\Sigma}$, and we make use of this index to introduce three related instances:
\begin{itemize}
    \item {\em Small time intervals}: ${\cal I}_-$ is the restriction of ${\cal I}$ to the set of items $S_- = \bigcup_{m < \hat{m}} S_m$. 
    
    \item {\em Large time intervals}: ${\cal I}_+$ is its restriction to $S_+ = \bigcup_{m > \hat{m}} S_m$. 

    \item {\em Merged instance}: ${\cal I}_{\pm}$ is the instance  obtained by unifying the item sets $S_-$ and $S_+$.
\end{itemize}

\paragraph{The Near-Additivity Theorem.} Similarly to Section~\ref{subsec:near_additive_nested}, we proceed by arguing that the best-possible peak inventory level for ${\cal I}_{\pm}$ cannot be much smaller than the sum of these measures for ${\cal I}_-$ and ${\cal I}_+$.

\begin{theorem} \label{thm:additivity_coprime}
$\opt^{ \mycont }( {\cal I}_{\pm} ) \geq \opt^{ \mycont }( {\cal I}_- ) + (1 - 2\eps) \cdot \opt^{ \mycont }( {\cal I}_+ )$.
\end{theorem}
\begin{proof}
The inequality in question is an immediate consequence of Lemma~\ref{lem:eps_shorter_additive}, as our construction guarantees that ${\cal I}_-$ is $\eps$-shorter than ${\cal I}_+$. To verify this claim, note that 
\begin{eqnarray}
\lcm( \{ T_i \}_{i \in S_-}) & = &  \prod_{i \in S_-} T_i \label{eqn:additivity_coprime_1} \\
& \leq & \Psi_{\hat{m}-1}^{ \pi( \Psi_{\hat{m}-1} ) } \label{eqn:additivity_coprime_2} \\
& \leq & \Psi_{\hat{m}-1}^{ 1.2551 \cdot \frac{ \Psi_{\hat{m}-1} }{ \ln \Psi_{\hat{m}-1} } } \label{eqn:additivity_coprime_3} \\
& = & 4^{ 1.2551 \cdot \frac{ \Psi_{\hat{m}-1} }{ \ln 4 } } \nonumber \\
& \leq & \frac{ 4^{ \Psi_{\hat{m}-1} } }{ \Psi_{\hat{m}-1} } \nonumber \\
& \leq & \eps \cdot \Psi_{\hat{m}} \nonumber \\
& \leq & \eps \cdot \min_{i \in S_+} T_i \ . \label{eqn:additivity_coprime_4}
\end{eqnarray}
Here, equality~\eqref{eqn:additivity_coprime_1} holds since $\{ T_i \}_{i \in S_-}$ are pairwise coprime. Inequality~\eqref{eqn:additivity_coprime_2} is obtained by observing that all time intervals in $S_-$ are upper-bounded by $\Psi_{\hat{m}-1}$. In addition, there could be at most $\pi( \Psi_{\hat{m}-1} )$ such values greater than $1$, where $\pi(\cdot)$ is the prime-counting function. Inequality~\eqref{eqn:additivity_coprime_3} follows by employing the upper bound $\pi(x) \leq 1.2551 \cdot \frac{ x }{ \ln x }$ for $x > 1$, due to \citet[Cor.~5.2]{Dusart18}. Finally, inequality~\eqref{eqn:additivity_coprime_4} holds since $\{ T_i \}_{i \in S_+}$ are lower-bounded by $\Psi_{\hat{m}}$.
\end{proof}

\subsection{Algorithmic implications}

Given this result, we begin by computing a $(1+\eps)$-approximate shift vector $\tau^-$ with respect to ${\cal I}_-$. For this purpose, since $T_i < \Psi_{\hat{m}} \leq \Psi_{ \frac{ 1 }{ \eps} + 1 }$ for every item $i \in S_-$, the number of distinct time intervals across this set is only $O( \Psi_{ \frac{ 1 }{ \eps} + 1 })$. Therefore, as argued in  Theorem~\ref{thm:main_result_K}, we can compute $\tau^-$ in time
\[ O\left( 2^{ \tilde{O}( \Psi_{ \frac{ 1 }{ \eps} + 1 } / \eps^2 ) } \cdot |{\cal I}|^{O(1)} \right) ~~=~~ O \left( \mytower_4 \left(O \left( \frac{ 1 }{ \eps} \right), O \left(\frac{ 1 }{ \eps} \right) \right) \cdot |{\cal I}|^{O(1)} \right) \ .  \]
For the instance ${\cal I}_+$, as well as for the set of items $S_{\hat{m}}$, we pick arbitrary shift vectors, $\tau^+$ and $\tau^{\hat{m}}$. Our final solution $\tau$ is formed by gluing $\tau^-$, $\tau^+$, and $\tau^{\hat{m}}$ together.

\paragraph{Large pairwise coprime time intervals.} While $\tau^-$ is a $(1+\eps)$-approximate shift vector with respect to ${\cal I}_-$, the performance guarantee of $\tau^+$ in terms of ${\cal I}_+$ is still unclear. Recalling that all time intervals of the latter instance are lower-bounded by $\Psi_{\hat{m}} \geq \frac{ 1 }{ \eps }$, the next claim shows that any shift vector is $(1+\eps)$-approximate in this context, since we obtain $\opt^{ \mycont }( {\cal I}_+ ) \geq (1 - \eps) \cdot H( S_+ )$ as a direct implication.

\begin{lemma}
Let ${\cal I}$ be a pairwise coprime inventory staggering instance. Then, 
\[ \opt^{ \mycont }( {\cal I} ) ~~\geq~~ \left( 1 - \frac{ 1 }{ T_{\min} } \right) \cdot H_{\Sigma} \ . \]
\end{lemma}
\begin{proof}
Our proof is based on arguing that $I_{\max}( \tau ) \geq ( 1 - \frac{ 1 }{ T_{\min} } ) \cdot H_{\Sigma}$, for every real-valued shift vector $\tau \in \bbR^n$. To this end, let $\hat{\tau} \in \bbZ^n$ be the vector specified by $\hat{\tau}_i = \lceil \tau_i \rceil$, for every item $i \in [n]$. Since the time intervals $T_1, \ldots, T_n$ are pairwise coprime, by the Chinese Remainder Theorem, the system of congruences
\[ t ~~\equiv~~ \hat{\tau}_i \ (\mymod \ T_i) \qquad \forall \, i \in [n] \]
has a unique solution $t$ modulo $\prod_{i \in [n]} T_i$. As such, for every $i \in [n]$, the policy ${\cal P}_{ \tau }$ has an $i$-order at time $t - (\hat{\tau}_i - \tau_i)$, implying that the inventory level of this item at time $t$ is
\[ I_i( { \tau }_i, t ) ~~\geq~~ H_i \cdot \left( 1 - \frac{ \hat{\tau}_i - \tau_i }{ T_i } \right) ~~=~~ H_i \cdot \left( 1 - \frac{ \lceil \tau_i \rceil - \tau_i }{ T_i } \right) ~~\geq~~ H_i \cdot \left( 1 - \frac{ 1 }{ T_{\min} } \right) \ . \]
It follows that the maximum inventory level of the policy ${\cal P}_{\tau}$ is $I_{\max}( \tau ) \geq I_{\Sigma}( \tau, t ) \geq ( 1 - \frac{ 1 }{ T_{\min} } ) \cdot H_{\Sigma}$.
\end{proof}

\paragraph{Final performance guarantee.} We conclude the proof of Theorem~\ref{thm:main_result_coprime}
by observing that 
\begin{eqnarray}
I_{\max}( \tau ) & \leq & I_{\max}^{ S_-} ( \tau^- ) + I_{\max}^{ S_+}( \tau^+ ) +  I_{\max}^{ S_{\hat{m}}}( \tau^{\hat{m}} ) \nonumber \\
& \leq & (1 + \eps) \cdot \opt^{ \mycont }( {\cal I}_- ) + H( S_+ ) + H( S_{\hat{m}} ) \label{eqn:comprime_final_1} \\
& \leq & (1 + 2\eps) \cdot \left( \opt^{ \mycont }( {\cal I}_- ) + \opt^{ \mycont }( {\cal I}_+ ) \right) + \eps H_{\Sigma} \label{eqn:comprime_final_2}  \\
& \leq & (1 + 2\eps)(1 + 3\eps) \cdot \opt^{ \mycont }( {\cal I}_{\pm} ) + 2\eps \cdot \opt^{ \mycont }( {\cal I} )  \label{eqn:comprime_final_3} \\
& \leq & ( 1 + 13\eps ) \cdot \opt^{ \mycont }( {\cal I} ) \ . \nonumber
\end{eqnarray}
Here, inequality~\eqref{eqn:comprime_final_1} holds since $\tau^-$ is a $(1+\eps)$-approximate shift vector with respect to ${\cal I}_-$. Inequality~\eqref{eqn:comprime_final_2} is obtained by recalling that $\opt^{ \mycont }( {\cal I}_+ ) \geq (1 - \eps) \cdot H( S_+ )$ and $H( S_{\hat{m}} ) \leq \eps H_{\Sigma}$. Finally, inequality~\eqref{eqn:comprime_final_3} follows from Theorem~\ref{thm:additivity_coprime} and Lemma~\ref{lem:avg_space_LB}.

%% file: TEX-Conclusions.tex
\section{Concluding Remarks}

We conclude our work by briefly discussing two pivotal questions that lie at the heart of inventory staggering. The first question is concerned with the long-standing approximability gap of this problem in its full generality, whereas the second touches upon peak inventory evaluation, which has been circumvented in earlier papers by focusing on stylized formulations.

\paragraph{Sub-$\bs{2}$-approximation/APX-hardness for general instances?} As explained in Section~\ref{subsec:prev_new_results}, the classic average-space  bound (Lemma~\ref{lem:avg_space_LB}) informs us that the peak inventory level of any shift vector is within factor $2$ of optimal, regardless of how a given inventory staggering instance is structured. In the opposite direction, this problem has been shown to be strongly NP-hard; see \cite{GallegoSS92} and \cite{HochbaumR19} for intractability proofs applying to the continuous and discrete settings, respectively. Without any simplifying assumptions, we believe that meaningfully narrowing the approximability region $[\mathrm{PTAS},2]$ would be a very challenging avenue for future research, probably requiring technical developments that are  unavailable within this domain at present time. 

\paragraph{Algorithmic/hardness results for peak evaluation?} As mentioned in Section~\ref{subsec:prev_new_results}, efficiently evaluating the peak inventory level attained by a given shift vector remains a central question to be addressed. Toward this objective, Theorem~\ref{thm:LB_random_sample} excludes standard sampling ideas, showing that $2^{ \Omega(\eps^2 n ) }$ random points should be drawn in order to estimate peak inventory levels within factor $\frac{ 1 }{ 2 } + \eps$ with constant probability. Along these lines, it would be interesting to examine whether exact/approximate peak evaluation is actually \#P-hard. On a brighter note, Theorem~\ref{thm:compute_peak} presents an exact evaluation oracle in $O ( n^{ O(n) } \cdot | {\cal I} |^{ O(1) } )$ time, by means of integer linear programming in fixed dimension \citep{Lenstra83, Kannan87, FrankT87}. Substantial improvements on the latter running time, as well as faster approximate procedures in this context, would be challenging directions to pursue.